\documentclass{aastex}
\usepackage[twocolumn]{emulateapj5}

\slugcomment{Astron. J., accepted 2007 Jan 18}

\def\ltsima{$\; \buildrel < \over \sim \;$}
\def\simlt{\lower.5ex\hbox{\ltsima}} 
\def\gtsima{$\; \buildrel > \over \sim \;$}
\def\simgt{\lower.5ex\hbox{\gtsima}} 
\def\arcsec{\hbox{$^{\prime\prime}$}}
\def\deg{\hbox{$^\circ$}}
\def\ch{\checkmark}
\usepackage{rotating}

\shorttitle{Candidate X-shaped Radio Sources} 
\shortauthors{Cheung}

\begin{document}

\title{FIRST `Winged' and `X'-shaped Radio Source Candidates}

\author{C.~C. Cheung\altaffilmark{1}}
\affil{National Radio Astronomy Observatory; and \\ 
Kavli Institute for Particle Astrophysics and Cosmology, Stanford
University, Stanford CA 94305}

\email{teddy3c@stanford.edu}
\altaffiltext{1}{Jansky Postdoctoral Fellow}

\begin{abstract}

A small number of double-lobed radio galaxies (17 from our own census of the
literature) show an additional pair of low surface brightness `wings', thus
forming an overall `X'-shaped appearance.  The origin of the wings in
these radio sources is unclear. They may be the result of back-flowing plasma
from the currently active radio lobes into an asymmetric medium surrounding the
active nucleus, which would make these ideal systems in which to study
thermal/non-thermal plasma interactions in extragalactic radio sources. Another
possibility is that the wings are the aging radio lobes left over after a
(rapid) realignment of the central supermassive black-hole/accretion disk system
due perhaps to a merger.  Generally, these models are not well tested; with the
small number of known examples, previous works focused on detailed case studies
of selected sources with little attempt at a systematic study of a large sample. 
Using the VLA-FIRST survey database, we are compiling a large sample of winged
and X-shaped radio sources for such studies. As a first step toward this goal,
an initial sample of 100 new candidate objects of this type are presented in
this paper.  The search process is described, optical identifications from
available literature data, and basic radio data are presented.  From the limited
resolution FIRST images ($\sim$5\arcsec), we can already confidently classify a
sufficient number of these objects as having the characteristic wing lengths
$>$80$\%$ of the active lobes to more than double the number of known X-shaped
radio sources.  We have also included as candidates, radio sources with shorter
wings ($<$80$\%$ wing to lobe length ratios), or simply `winged' sources, as it
is probable that projection effects are important. Finally, among the candidates
are four quasars ($z$=0.37 to 0.84), and several have morphologies suggestive of
Fanaroff-Riley type-I (low-power) radio galaxies. While followup observations
are necessary to confirm these identifications, this stresses the importance of
source orientation and imaging limitations in finding these enigmatic objects. 

\end{abstract}

\keywords{Galaxies: active --- galaxies: jets --- quasars: general ---
radio continuum: galaxies}

\section{Background and Motivation}

\begin{figure}
\epsscale{1}
\plotone{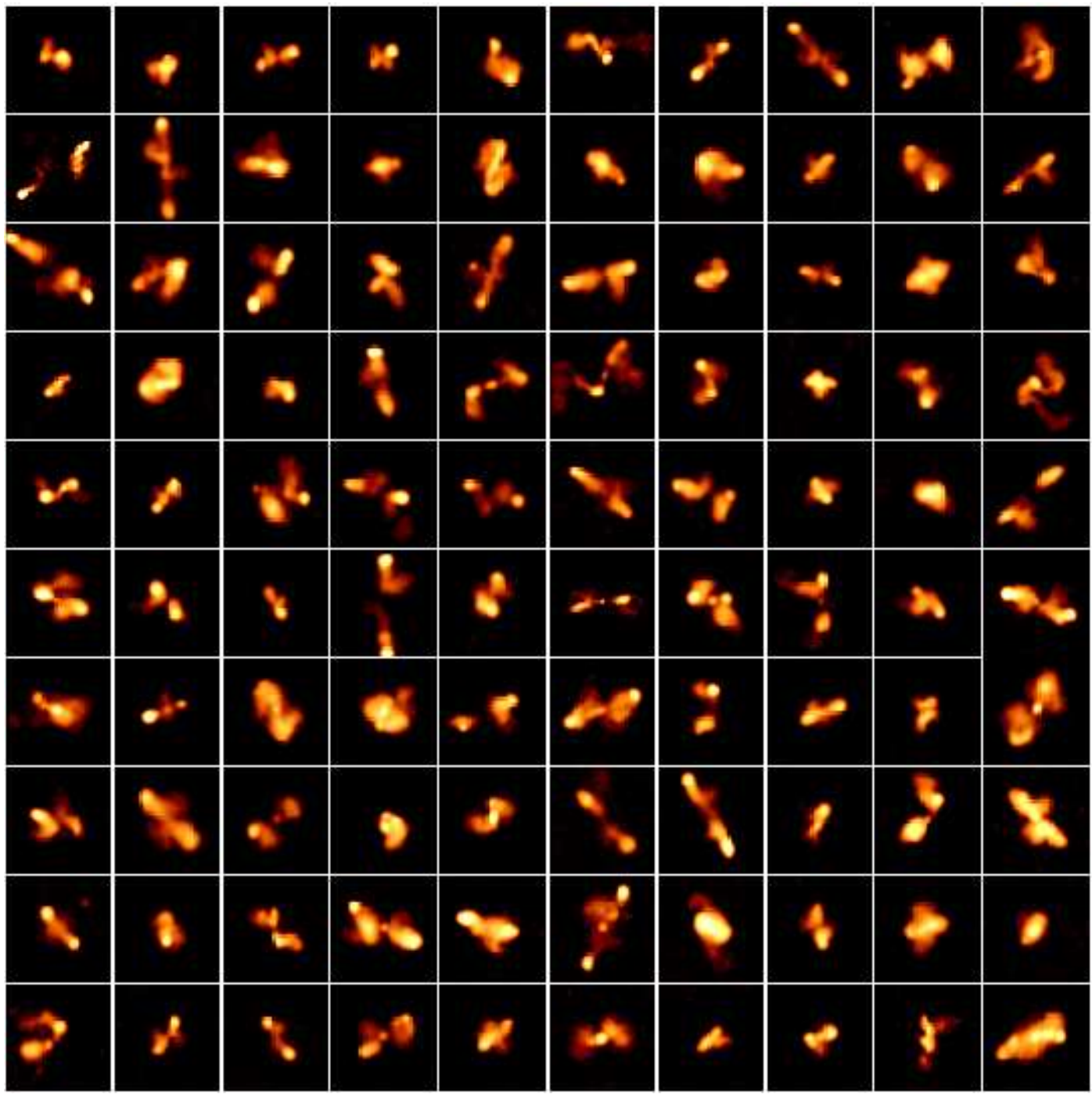}
\figcaption[f1.eps]{\label{fig-1}
Color thumbnail images of the 100 FIRST X-shaped radio source candidates.
The objects are ordered in increasing R.A. order (refer to the catalog 
numbers in Table~\ref{tbl-new}) from left to right then down within each 
of the four 5$\times$5 panels (top left = cat. no. 1--25, top right = 
cat. no. 26--50, bottom left = cat no. 51--75, bottom right = cat. no. 
76--100).  See Figure~\ref{fig-new} for image scales.}
\end{figure}

`Winged', or `X'-shaped radio galaxies form a small and interesting
subclass of extragalactic radio sources \citep[e.g.,][]{lea92,den02}. In
addition to the usual pair of `active' lobes, these objects
possess an additional pair of lower surface brightness `wings' of
emission, thus forming an overall winged or X-shaped appearance (see
Figure~\ref{fig-1} for examples). 

There is no clear explanation for the origin of these unusual
morphologies. One possibility is that the wings result from back-flow
\citep{lea84} of plasma from the ``working surfaces'' (hot spots) in the
active lobes into an asymmetric surrounding medium, with subsequent
buoyant expansion \citep{wor95}. In this scenario, it has been argued that
the expansion of the wings is subsonic and it becomes difficult to explain
sources with wings that are longer than the active lobes \citep{den02}.
However, this supposes that the active lobes are expanding supersonically,
and into a roughly spherically symmetric medium -- neither assumption may
be applicable to these objects \citep{cap02,kra05}. 

Another explanation that has received much attention lately is that the
wings are the remnant left over from a rapid realignment of a central
supermassive black hole/accretion disk system \citep[e.g.,][]{den02},
perhaps due to a relatively recent merger of a binary supermassive
black-hole (SMBH) system \citep[e.g.,][]{rot01,mer02,gop03,zie05}.  The
X-sources are then one of only a few observational systems attributable to
SMBH binaries \citep{kom03}. In this respect, a census of these sources
is important to estimate the occurrence rate of mergers \citep[e.g.,
][]{mer02,hug03} and recurrence time-scales of (radio-loud) active 
galaxies \citep[e.g.,][]{liu04}.

One major obstacle in understanding the origin of the X-shaped radio sources as
a class is that there are only a small number of known examples -- 17 in total
from our own census of the literature (Table~\ref{tbl-known}), though wings in
radio galaxies are fairly common (\S~\ref{section-def}).  It is unclear whether
these known ones are representative of the class as a whole. They are in the
majority, identified in imaging surveys of radio samples with high flux density
thresholds (\S~\ref{section-radiolum}). This may explain their limited range of
properties: the well-known examples are relatively local (within z$<$0.3), and
the radio luminosities straddle the Fanaroff-Riley (FR) division \citep{fan74},
with a dearth of sources showing FR-I morphology \citep{den02}. Of these, only a
handful have been studied in sufficient detail to compare to the differing model
expectations \citep{wor95,ulr96,den02,kra05}. Until only recently
\citep[\S~\ref{section-census};][]{wan03,lan06}, none showed broad-emission
lines characteristic of quasars (i.e., the population of radio galaxies aligned
close to our line of sight), stressing the importance of source orientation in
identifying new examples (\S~\ref{section-appearance}). 

We can begin to address these issues by using larger radio imaging surveys
to search for more examples of X-sources to expand the current sample for
follow-up studies. We are carrying out such a project in search of new
winged and X-shaped radio sources using the FIRST (Faint Images of the Radio Sky at
Twenty-centimeters) survey \citep{bec95} (\S~\ref{sec-id}).  Our simple
search procedure, described here, and applied to a subset of the FIRST
database, resulted so far in 100 radio sources we propose as candidates
(\S\S~\ref{sec-id}~and~\ref{sec-notes}).  Although there are already a
sufficient number of clear new examples of bona-fide X-shaped radio sources to at least double the known
sample, the majority require higher resolution and better sensitivity
observations to map out their morphologies more clearly
(Figure~\ref{fig-1}). We have included sources with shorter wings
\citep[canonically, $<$80$\%$ lobe length;][]{lea92} with the hopes of
disentangling projection effects. As part of the search process, we have
compiled an up-to-date catalog of known X-shaped radio sources from the
literature which is used to evaluate the success of our search methods
(\S~\ref{section-known}). A summary is given in \S~\ref{section-summary}. 

Throughout this paper, we assume a $\Lambda$CDM cosmology with parameters
$h=H_{0}$/(100 km~s$^{-1}$ Mpc$^{-1})=0.7$, $\Omega_{\rm M}=0.3$ and 
$\Omega_{\rm \Lambda}=0.7$.

\section{Definition and Census}\label{section-def}

As part of our search for new examples, we compiled a list of winged and
X-shaped radio sources commonly discussed in the literature
(Table~\ref{tbl-known}). Fourteen well-known examples are listed by
\citet{rot01}, \citet{cap02}, and \citet{mer02} (attributed to \citet{lea92}). 
The X-shaped nature of the radio source 4C+04.40 \citep{jun00} was revealed
contemporaneously to these works. \citet{wan03} recently reported 4C+01.30 as
the first X-shaped source with a quasar nucleus (i.e the population of radio
galaxies with the main axis aligned closer to our line-of-sight), showing broad
emission lines in its spectrum.  One more quasar has since been identified as an
X-shaped radio source \citep{lan06}, and we located two additional, more distant
\citep[$z\sim$0.4;][]{leh01,zak04} X-shaped radio galaxies in published maps.
See the Appendix for notes on these five additional examples from the
literature. 

The known examples show a range of wing to lobe length ratios (in projection).
This brings up the issue of nomenclature and the characteristics distinguishing
an X-shaped source from a winged one. Conventionally, winged radio sources with
the most extended wings \citep[$>$80$\%$ of the extent of the lobes;][]{lea92}
are called X-shaped, but these are clearly phenomenologically related to shorter
winged objects. The radio galaxy 4C+32.25 shows the most prominent wings, being
just over a factor of two more extended than its lobes.  3C~192 and 3C~379.1
have the shortest wings \citep[$\sim$65$\%$ of extent of
lobes;][respectively]{den99,spa85}, so by convention, do not qualify as being 
X-shaped. However, the wings in the latter two examples may appear shortened by
projection, or they may be caught in a stage of early growth or later decay (see
\S~\ref{section-appearance}).  It is apparent that such effects should be
accommodated in further discussions of these objects as a class. 

The present search aims to identify new X-shaped radio sources, and winged ones
are included with the points above in mind.  In fact, wings are quite commonly
seen in deep radio images of 3CR radio galaxies \citep[e.g.,][]{lea84,lea91},
but we will not attempt to catalog them all here.  Deeper observations of our
candidates should determine the true extent of the wings where they may not be
so apparent in the limited sensitivity FIRST survey finder images.  For these
reasons, we use winged and X-shaped interchangeably throughout the rest of the
text.

\section{Identifying New Winged and X-shaped Radio Source Candidates with 
FIRST} \label{sec-id}

The FIRST survey uses the NRAO\footnote{The National Radio Astronomy
Observatory is operated by Associated Universities, Inc. under a
cooperative agreement with the National Science Foundation.} Very Large
Array (VLA) at a wavelength of 20cm (1.4 GHz) to image the radio sky with
an angular resolution of $\sim$5\arcsec\ \citep{bec95}.  This relatively
high resolution and the good sensitivity achieved in the survey images
(typical rms of 0.13 mJy) has allowed investigations of new fainter
samples of morphologically distinct radio sources such as, compact steep
spectrum \citep[CSS;][]{kun02}, hybrid morphology \citep[with FR-I and II
characteristics;][]{gaw06}, and core-dominated triples \citep{mar06}.
Here, we are similarly exploiting this unique database to search for new
examples of X-shaped radio sources. 

One of the main survey products is a catalog of radio ``sources'' derived
from fitting 2-dimensional Gaussians (with deconvolved major and minor
axes, $maj.$ and $min.$, respectively, and fitted peak flux) to the
images. By virtue of the high angular resolution of the survey, many of
these sources are often the separate components in a single object
decomposed, e.g. the lobes and/or hot spots in a double-lobed radio
galaxy. Most are simply unresolved point sources, or attempts by the
software to fit more complicated source structures with a single Gaussian
when they were inadequately resolved -- see \citet{pro03,pro06} for a
discussion. 

This catalog of radio sources can be queried through the search engine on
the FIRST website (http://sundog.stsci.edu/). We ignored sources flagged
by the software as sidelobes due to a nearby bright source. The latest
data release (2003 April
11)\footnote{http://sundog.stsci.edu/first/catalogs/readme$\_$03apr11.html\label{foot-first}}
was utilized;  the sky coverage in this version of the catalog is
approximately 8422 deg$^{2}$ in the north galactic cap (7.0
hr$<$R.A.$<$17.5 hr, $-$8.0 deg$<$Dec.$<$+57.6 deg), and 611 deg$^{2}$ in
the south galactic cap (21.3 hr$<$R.A.$<$3.3 hr, $-$11.5 deg$<$Dec.$<$+1.6
deg). The current FIRST catalog contains $\sim$811,000 sources
(footnote~\ref{foot-first}). 

\subsection{FIRST Search Procedure\label{section-search}}

With the large number of images fields provided by FIRST, our strategy is to
visually examine only a well-defined subset of these: 

\begin{itemize}

\item fields with sufficient dynamic range (defined as the ratio of the
peak : rms) in the images to be able to see extended low surface
brightness wings, and

\item those containing sources with resolved structure easily discernible
with the $\sim$5\arcsec\ resolution of the survey. 

\end{itemize}

\noindent These two issues are discussed in turn.

With a given image sensitivity, it becomes increasingly difficult to
distinguish the morphology of a low image peaked radio source by-eye
(below a few mJy/bm in FIRST).  For X-shaped sources, the surface
brightness of the wings are typically much lower than the brighter active
lobes/hot spots (the usual brightest parts of a radio galaxy). 
Quantitatively, low-peaked fields of X-sources yield at most, only a few
sigma detections of possible wing emission.  In this work, we determined
that fields with image peaks of 5 mJy/bm and greater ($\sim$40:1 dynamic
range) worked reasonably well enough (\S~\ref{section-census}) to see the
extended emission and set this as the limit in our queries. 

Ideally, we would have liked to inspect every field containing at least
one resolved source indicating the presence of an extended lobe. This can
be done by setting both $maj.$ and $min.$$>$5\arcsec, the resolution of
the FIRST images. However, setting this along with the $>$5 mJy/bm image
peak limit returned a large number of sources (8,146), with a small
percentage in overlapping fields whenever more than one component in the
field matched the search criteria. (For reference, setting a twice fainter
peak level returned approximately double the fields in our queries.) For
this initial phase of our project, we relaxed the fitted major axis to
$maj.$$>$15\arcsec\ which returned a much more manageable 1,648 source
fields for visual inspection. We hope to address the remaining fields in
future iterations. 

Our queries returned color GIF images of the 5$\times$5 arcmin$^{2}$ FIRST
fields which we inspected by-eye and downloaded the FITS images of the
fields with hints of extended structure characteristic of wings. The FITS
images were then further scrutinized using the SAOImage-ds9 interface to
control the contrast in the images. A working list of candidates was
generated for optical identifications with rough positions estimated from
the radio morphologies, usually where the radio lobes intersect when a
central source was not present. The final list of candidates were judged to
show at least one wing roughly orthogonal to a pair of active lobes.

\subsection{Optical Identifications and Properties}

We cross-referenced our radio-based estimated positions with
NED\footnote{This research has made use of the NASA/IPAC Extragalactic
Database which is operated by the Jet Propulsion Laboratory, California
Institute of Technology, under contract with the NASA.}, the Digital Sky
Survey (DSS) based USNO-B1.0\footnote{This research has made use of the
USNOFS Image and Catalogue Archive operated by the United States Naval
Observatory, Flagstaff Station (http://www.nofs.navy.mil/data/fchpix/);
the ``-B1.0'' is suppressed hereafter.} \citep{mon03}, and APM
\citep{mad90,mcm02} catalogues, and the Sloan Digital Sky
Survey\footnote{Funding for the SDSS and SDSS-II has been provided by the
Alfred P. Sloan Foundation, the Participating Institutions, the National
Science Foundation, the U.S.  Department of Energy, the National
Aeronautics and Space Administration, the Japanese Monbukagakusho, the Max
Planck Society, and the Higher Education Funding Council for England. The
SDSS Web Site is http://www.sdss.org/.} (SDSS). The identifications were
based on the position of the optical object relative to the radio source
morphology.  This was aided by making radio/optical overlays of the fields
using the FIRST and red-filter DSS images (Figure~\ref{fig-new}).  Known
X-shaped sources (\S~\ref{section-known}) were identified and removed 
from the candidate list. 

For the objects without catalogued or visible optical counterparts (6
cases), we kept the best guess radio-morphology based positions; these are
less precise because the radio emission is extended.  An (uncatalogued)
smudge appears near the radio-predicted position of J1655+4551 in the DSS
red image -- we made a by-eye position measurement of this optical source. 

Of the 100 candidates, 36 are spectroscopically identified, with half of
them (18/36) found in the latest SDSS data release \citep[DR5;][]{ade07}. This is
no surprise, as the FIRST survey was designed to have overlapping coverage
with the SDSS.  One new identification was obtained for us by S.E. Healey
(2005, private communication) observing at the MacDonald Observatory
(J0702+5002; see notes in \S~\ref{sec-notes}). The remaining spectroscopic
identifications are from prior work found in the literature; three of
these have SDSS data confirming their IDs.  The positions, optical (SDSS
or POSS-II) $r$-band magnitudes, $g-r$ colors of the SDSS identified
objects, and spectroscopic identifications \& redshifts for the 100 final
candidates are tabulated in Table~\ref{tbl-new}. 

Among the 36 spectroscopically identified candidates are four quasars;  the
remainder are radio galaxies. The quasars were identified by the SDSS as having
at least one broad ($>$1000 km s$^{-1}$) emission line \citep{sch02}. The three
highest redshift (z$>$0.5) quasars have bluer colors ($g-r$=--0.1 to 0.2),
indicative of a quasar nucleus dominating the optical continuum.  The fourth
quasar (J1430+5217), at a more modest redshift of $z$=0.367, has the clearest
winged appearance among the quartet (Figure~1), although curiously has a redder
color ($g-r$=1.1) more like the radio galaxies.  The unidentified J1406+0657
($g-r$=0.0) and to a lesser degree, J1342+2547 ($g-r$=0.4), are bluer like the
highest redshift quasars; spectroscopic observations are necessary to determine
if they are quasars also.  Other higher redshift candidates of note are the
radio galaxies J0143--0119 ($z$=0.52), J0115--0000, J0941--0143, and J1309--0012
(latter three are at $z$$\sim$0.4). 

Figure~\ref{fig-new} shows the fields of the 100 candidate X-shaped
sources in the radio (FIRST) and the optical (Digital Sky Survey red
images) centered on the best determined positions.  Each field is
2$\times$2 arcmin$^{2}$, except for the cases of the two largest angular
size radio sources (J0113+0106 and J1424+2627) where larger 4$\times$4
arcmin$^{2}$ fields are displayed. Thumbnail color versions of the FIRST
images which emphasize the diffuse emission are shown in
Figure~\ref{fig-1}. Notes on each candidate are provided in
\S~\ref{sec-notes}.

\subsection{Basic Radio Properties\label{sec-radio}}

We have compiled some basic radio data for the 100 X-shaped source
candidates: when available, radio flux densities at 365 MHz from the TXS
survey \citep{dou96}, 1.4 GHz from the NVSS \citep{con98}, and at 4.9 GHz
from the PMN \citep{gri94,gri95}, Green Bank \citep{bec91,gre91}, and
Parkes \citep[PKS 5 GHz;][]{wri90} surveys. These radio data are tabulated
in Table~\ref{tbl-new}. 

NVSS 1.4 GHz measurements are available for every object (as opposed to
the other frequencies) and are probably the most reliable {\it total}
radio flux densities for our candidates. It is the most sensitive, and has
the best resolution ($\sim$45\arcsec) compared to the other frequency
surveys, so is least susceptible to source confusion. The lower resolution
VLA configuration used for the survey (compared to the FIRST) means it is
better suited to detecting the most extended radio structure. 

In six of the largest radio galaxies (J0113+0106, J0147--0851, J0943+2834,
J1424+2637, J1501+0752, J1606+0000), it is apparent that they are doubles
in the NVSS maps and the integrated fluxes were measured by us. In the   
case of J1327--0203, we measured the integrated 1.4 GHz flux density from
the FIRST map, rather than the NVSS one, so that the 60 mJy contaminating
source at the northern edge of the 2' field could be accounted for
(Figure~\ref{fig-new}).

We also inspected the NVSS maps to search for possible larger scale radio
structure and to identify potential misidentifications (e.g. our candidate
may be part of a field source of another object; see
\S~\ref{section-census}). Only two objects, J1049+4422 and J1339--0016,
showed hints of diffuse emission that may be associated with our radio
source (i.e. they are not due to additional point sources in the field as
seen in the higher resolution FIRST images). These possible associations
are difficult to determine conclusively because of the mismatch in
resolution between the maps (factor of 10 difference between NVSS and
FIRST) so this is just mentioned for completeness. 

A number of the TXS positions were offset from our determined ones (by
$\simgt$1' in some cases).  This is because of the extended (and
asymmetric) source structure of many of these radio sources and the
limited resolution of the TXS measurements. In the case of the faint
source J1456+2542 (36 mJy from NVSS), it is apparent that the TXS
measurement was contaminated by other sources visible in the NVSS field so
was discarded. 

Two-point radio spectral indices relative to 1.4 GHz are also calculated
when data at one or both of the other two frequencies were available. No
such information was available for 6/100 sources which are among the
faintest at 1.4 GHz. Of the remaining, most (90/94) show steep radio
spectra ($\alpha$$\geq$0.5; $F_{\nu}\propto\nu^{-\alpha}$) between at
least one frequency pair as expected from lobe-dominated radio sources.
The remaining four are among the faintest sources, and have single
spectral index values as small as $\sim$0.3; we consider these consistent
with a steep spectrum considering the possible uncertainties in such low
flux density measurements.

\section{Notes on the Candidate X-shaped Radio Sources}
\label{sec-notes}

Here, we provide notes on the optical fields and our interpretation of the
radio morphologies for each X-shaped radio source candidate in the order 
presented in Table~\ref{tbl-new}. 

\smallskip\noindent{\bf (1) J0001--0033:} This is a double (east-west) radio
source with a clear optical counterpart positioned close to the western
lobe. The galaxy is spectroscopically identified by the SDSS with a
redshift of 0.25. The more prominent wing is north of the (fainter)
eastern lobe in the FIRST image. 

\smallskip\noindent{\bf (2) J0033--0149:} The radio source looks like it
could be a winged low-power centrally bright (FR-I) radio galaxy except
that we have chosen to associate it with the brightest optically
identified source toward the northern edge of the radio source. The bright
optical galaxy is at $z$=0.13 \citep{jon05}. 

\smallskip\noindent{\bf (3) J0036+0048:} There is a faint optical object in
between the active radio lobes which are approximately oriented east-west.
The shorter of the two wings is south of the western lobe. 

\smallskip\noindent{\bf (4) J0045+0021:} Early higher resolution radio
maps of this radio source, a.k.a. 4C~--00.05 \citep{dow86,jac99} actually
already show the clear X-shaped morphology although it was never noted by
these authors.  \citet{dun89} identified a faint optical counterpart
($r$=21.5 mag.) without a redshift determination -- it is too faint to see
clearly in the DSS plates. The two most prominent galaxies in the 2' field
are probably physically associated with each other, although their
relationship to 4C~--00.05 is unclear: SDSS J004539.67+002057.6 at
$z$=0.116 (to the west) and SDSS J004543.18+002145.9 at $z$=0.115 (to the
north). 

\smallskip\noindent{\bf (5) J0049+0059:} The optical source is closer to the
northern lobe than the southern one and there is diffuse emission in both
the east and west directions which may be wings. The optical
identification as a $z$=0.30 galaxy is via the SDSS database.
Additionally, there is a prominent $z$=0.106 galaxy \citep{rin03} $\sim$1'
to the northeast (Figure~\ref{fig-new}). 

\smallskip\noindent{\bf (6) J0113+0106:} This is one of the clearest new
examples of an X-shaped source, mainly because of its larger angular
extent in comparison to our other candidates. An available lower
resolution \citep[NVSS;][]{con98} map (not shown) shows the faint wing
emission more clearly. The optical counterpart \citep[$z$=0.281;][]{lac00}
is clearly coincident with a faint radio core peaked at 1.4 mJy/bm in the
FIRST map.  \citet{got02} used SDSS data to identify a cluster of galaxies
about 1.5' away directly to the north, but at an estimated $z$=0.186. 

\smallskip\noindent{\bf (7) J0115--0000:} The optical counterpart to this
radio galaxy is a faint source in the DSS red plate at $z$=0.38
\citep{lac00}. The active radio lobes (north-south) are rather symmetric
with clear (although short) wings in the east-west directions. 

\smallskip\noindent{\bf (8) J0143--0019:} This high redshift ($z$=0.52) radio
galaxy was identified by \citet{lac00}. From the position he provided
[J2000: R.A.=01h43m17s, Dec.=--01d18m59s], we infer that the optical
counterpart is the faint smudge in the DSS plate in the eastern lobe
(catalogued by the USNO as having a similar red magnitude as provided by
\citet{lac00}). If this is indeed the case, then the radio source has
quite an asymmetric appearance. The position we provide is based roughly
on where the east and west lobes intersect -- a higher resolution 5 GHz
image \citep{rei99} shows a bright, compact radio feature near this
location. The northern wing is very prominent, but the southern one is
not. 

\smallskip\noindent{\bf (9) J0144--0830:} The faint radio source is centrally
peaked like an FR-I radio galaxy in the low resolution FIRST image. An
optical counterpart near the radio peak is found in the USNO and SDSS
catalogues. We imagine the main active radio axis is roughly east-west
(more extended) with wings in the orthogonal direction. 

\smallskip\noindent{\bf (10) J0145--0159:} The optical counterpart is the
brighter, more centrally located object (relative to the radio source) in
the DSS plate. This brighter optical source is identified to be at
$z$=0.126 \citep{jon05} but the fainter (northern) one is unidentified.
The (north-south) radio source has clear edges seen to the NW and SE with
the wings opposite of them.  The $b$=19 mag object to the south [J2000:
R.A.=01h45m19.99s, Dec.=--02d00m24.7s] may have a radio counterpart peaked
at 0.4 mJy/bm. 


\smallskip\noindent{\bf (11) J0147--0851:} A faint optical source in the USNO
and SDSS catalogs is clearly visible on the DSS red plate and centrally
located with respect to the double (NE-SW) radio source. The wing to the
south is the more prominent one. 

\smallskip\noindent{\bf (12) J0211--0920:} A pair of optical sources
4\arcsec\ apart are found near the center of the radio source. We
associate the northeastern one of the pair as the optical counterpart
since it relates slightly better to the overall radio structure. The
western lobe is the brighter one and northern wing is shorter. 

\smallskip\noindent{\bf (13) J0225--0738:} There is a very faint smudge 
on the DSS plate just to the southern edge of the northern lobe which is
catalogued by the SDSS with $r$=24.3 mag. The more prominent wing is east
of the northern lobe and there is some hint of a shorter wing west of the
southern lobe. 

\smallskip\noindent{\bf (14) J0702+5002:} The radio source shows very
prominent wings to the northeast and southwest. The morphology is
suggestive of a FR-I radio galaxy as it is not clearly edge brightened,
although this needs confirmation with higher resolution imaging. S.~E. 
Healey (2005, private communication) kindly obtained a spectrum from a
600s exposure at the 2.7m Harlan J.  Smith Telescope at the MacDonald
Observatory on Oct 30, 2005 and determined a galaxy redshift of 0.0946. 

\smallskip\noindent{\bf (15) J0725+5835:} A faint optical smudge from the
USNO catalog is found near the center of the radio galaxy.  There is a
hint of a small wing to the west. To the east, the possible winged
emission is confused with radio emission from a faint optical source with
a radio counterpart ($\sim$7 mJy/beam peak in the FIRST image). 

\smallskip\noindent{\bf (16) J0805+4854:} This is one of the smaller angular
size radio sources presented in this paper so it is difficult to identify
the possible winged emission. The hint of a wing to the southwest is the
reason we have included it. There is no clear optical counterpart to the
limit of the DSS red plate.  There is a very faint $r$=22.1 ($g-r$=1.1)
source SDSS J080543.58+485505.1 toward the northern edge of the radio
source which may be the counterpart. Also, a $g$=16.8 mag SDSS galaxy at
$z$=0.055 appears at the top edge (about 1' to the north) of the 2' wide
field of our overlays (Figure~\ref{fig-new}). 

\smallskip\noindent{\bf (17) J0813+4347:} The active radio lobes appear to
run east-west and the more prominent wing is to the north. The optical
galaxy is at a modest redshift of $z$=0.128 as identified by the SDSS. 

\smallskip\noindent{\bf (18) J0821+2922:} This radio galaxy would be very
asymmetric if based on the position and identification ($z$=0.25) given by
\citet{wil03}. Their position is derived from near-infrared imaging (there
is indeed a faint optical smudge in the DSS plate at this position with a
faint ($r$=20 mag) counterpart in the SDSS images -- we have centered the
field in Figure~\ref{fig-new} on this position). The radio source is
dominated by the southwest lobe in the FIRST image and the southeast wing
is very prominent. 

\smallskip\noindent{\bf (19) J0836+3125:} We have centered the field
(Figure~\ref{fig-new}) on a faint optical source appearing in the USNO and
SDSS catalogues ($r$=20 mag). A clear wing appears east of the brighter
northern lobe. 

\smallskip\noindent{\bf (20) J0838+3253:} This is one of our less certain
identifications as a winged source since it can simply be a slightly
distorted FR-I radio galaxy.  The peak in the FIRST radio image is offset
by about 2\arcsec\ just to the south of east of the optical galaxy
identified to be at $z$=0.21 by the SDSS. 

\smallskip\noindent{\bf (21) J0845+4031:} The optical counterpart is
centrally located in between the two (east-west) active lobes, and the
wings are quite prominent. There is a slight twist traced by the radio
emission running from the nucleus to the peaks in the lobes. The brighter
galaxy about 0.5' to the southwest may just be a field galaxy (identified
as a $g$=18.2 mag galaxy at $z$=0.096 by the SDSS). 

\smallskip\noindent{\bf (22) J0846+3956:} This is a double (SE-NW) radio
source with hints of wings to the east and west. There is no obvious
optical counterpart in the DSS image but the SDSS database does give a
$r$=20.8 mag. counterpart. 

\smallskip\noindent{\bf (23) J0859--0433:} This is a very clear example of
a winged radio galaxy. The main axis is roughly east-west with a bright
hot spot like source at the edge of the western lobe. There is a faint
unidentified optical counterpart catalogued in the USNO. 

\smallskip\noindent{\bf (24) J0914+1715:} The FIRST image shows an asymmetric
pair of wings with the more pronounced (southern) wing appearing to be
associated with the brighter western lobe. A higher resolution 5 GHz image
from \citet{haa05} shows the radio core and the two active lobes
characteristic of an FR-II, but the wings are not visible in their image.
The optical counterpart is just a faint smudge on the DSS plate and is
catalogued by the SDSS at $r$=20 mag. 

\smallskip\noindent{\bf (25) J0917+0523:} A symmetric double radio source
is seen in the FIRST image, with a faint smudge on the DSS plate and
catalogued in the SDSS as $r$=20 mag. A bright optical field source is
found at the northern edge of the eastern lobe. The wings are
approximately orthogonal to the axis (east-west) of the active lobes. 

\smallskip\noindent{\bf (26) J0924+4233:} This radio source is identified
with a $z$=0.23 galaxy by the SDSS. Wings are quite clearly extended to
the north and south of the galaxy. About 1.7' to the southeast (not shown
in Figure~\ref{fig-new}) is a $z$=0.17 quasar (SDSS J092451.40+423218.9)
also recorded by SDSS. 

\smallskip\noindent{\bf (27) J0941--0143:} There are bright symmetric
radio lobes aligned to the NE-SW in this $z$=0.38 galaxy associated with a
small cluster \citep{spi79}. While there is a hint of a short west wing,
the eastern one is more prominent. 

\smallskip\noindent{\bf (28) J0941+2147:} The peak in the southern radio lobe
is about 5.5 times larger than the northern one. There are two optical
sources in the line of sight of the (approximately north-south) radio
lobes, toward their edges: a $r$=19.4 mag one to the north and $r$=18.2 to
the south as catalogued in the USNO but neither appear to be the optical
counterparts. Instead, a very faint smudge ($r$=22.6) is catalogued by the
SDSS at the center of the double radio source. A sign of a wing is
apparent to the west. 

\smallskip\noindent{\bf (29) J0943+2834:} The double-lobed radio source has a
fairly large angular size with a shorter wing to the east associated with
the southern lobe. The faint optical source is near a faint central radio peak
just discernible in the FIRST image (0.6 mJy/bm). 

\smallskip\noindent{\bf (30) J1005+1154:} The active radio lobes run roughly
north-south with a clear optical counterpart in between them identified
by SDSS as a $z$=0.166 galaxy. The galaxy to the southeast about 40\arcsec\
distant has extended radio emission associated with it. 

\smallskip\noindent{\bf (31) J1008+0030:} This is a radio source associated
with the X-ray bright \citep{bri94} cluster Abell~0933 at $z$=0.098
\citep{owe95}. The optical and radio peaks are nearly coincident
\citep{owe92} being just 1-2\arcsec\ offset in our overlay. The FIRST map
shows the diffuse emission defining the proposed wings to the SE and NW
much better than in a previously published VLA map \citep{owe92}. 

\smallskip\noindent{\bf (32) J1015+5944:} This is a higher redshift object at
$z$=0.53, identified as a quasar by SDSS. It is also an X-ray source
\citep{zic03}. The eastern lobe is about 10 times brighter than the
eastern one suggesting that this is the direction of the fore-coming jet.
A wing is apparent to the north of the core coincident with an 8.5
mJy/beam peak in the radio. One hopes that the southern wing becomes
apparent with deeper observations. 

\smallskip\noindent{\bf (33) J1040+5056:} This radio source looks like a
low resolution image of a typical FR-I radio galaxy \citep[e.g.,
M84;][]{lai87}. The SDSS identifies the galaxy to be at $z$=0.154. The
brighter ($g$=18.2), more northerly of the two galaxies to the northwest
is identified at $z$=0.134 by the SDSS (J104017.65+505700.1). 

\smallskip\noindent{\bf (34) J1043+3131:} The radio source is associated with
the brightest central galaxy of the triple system \citep{fan77} at
z$\sim$0.036 \citep{fal99} and is an X-ray source \citep{wor00}. It has
been previously mapped with the VLA \citep[e.g.,][]{fan86,par86} showing
hot spots aligned at PA$\sim$162\deg. The extended emission perpendicular
to this axis can then be interpreted as wings. 

\smallskip\noindent{\bf (35) J1049+4422:} The faint optical counterpart is
closer to the northwest lobe, which shows the apparent wing to the south.
This galaxy is only about 1.5 arcmin from the galaxy cluster Abell 1101 at
$z$=0.23 which is to the southeast \citep{str99}. 

\smallskip\noindent{\bf (36) J1054+5521:} There is no clear optical
counterpart to the radio source in the DSS plates thus the position is
based on where we expect the optical counterpart to be relative to the
radio structure (in between the east-west lobes). Some faint optical
smudges appear toward the edges of the eastern lobe. 

\smallskip\noindent{\bf (37) J1055--0707:} The radio source is quite
symmetric looking centered about the bright optical counterpart. The DSS
image has a streak across it with unknown origin. The wings are quite
clear poking out the sides of the axis of the main (NE-SW) lobes. 

\smallskip\noindent{\bf (38) J1102+0250:} The SDSS cataloged a faint
($r$=22.3 mag) counterpart near the central radio peak (12 mJy/beam); this
is not obvious in the DSS plates. The could simply be a normal
double-lobed radio source but the diffuse structure to the southwest of
the southeast lobe is suggestive of winged emission. 

\smallskip\noindent{\bf (39) J1111+4050:} The radio source appears to be
associated with the $z$=0.074 galaxy MCG+07-23-030 in the X-ray bright
\citep{led03} cluster Abell~1190 \citep{sli98}. If this is the case, this
is unlikely to be an X-shaped radio source because of the implied
asymmetry; our original interpretation would have been that the active
lobes are marked by the roughly east-west peaks with a very prominent wing
to the south. 

\smallskip\noindent{\bf (40) J1114+2632:} A roughly centrally peaked radio
source (64 mJy/beam) in the FIRST map with a faint optical counterpart of
22.0 mag in the APM blue plate but without a red counterpart ($>$20.0 mag)
in the limit of the APM plate \citep{sne02}. These are consistent with the
SDSS detections (e.g. $r$=21.0 mag). The low frequency spectrum (between
0.365 and 1.4 GHz) is steep $\alpha$=0.77, then flattens to 5 GHz,
suggesting that an optically thick component emerges at the higher
frequencies. 

\smallskip\noindent{\bf (41) J1120+4354:} A fairly symmetric (NE-SW)
double-lobed radio source with small wings about a faint optical source in
the red DSS image ($r$=20.4 mag in the SDSS). The radio field source to
the northwest is peaked at 4.6 mJy/bm. 

\smallskip\noindent{\bf (42) J1128+1919:} The bright $r$=16.1 mag optical
source catalogued by the USNO matching the position of the peak in the
southern radio lobe (about 13\arcsec\ from the center of the radio source)
is probably a field source. Optical emission from the center of the radio
source is not visible in the DSS image so we base the position on the
radio structure.  Curiously, the SDSS catalogs an optical source very near
this approximated position (SDSS J112838.05+191956.7), however, there is
hardly anything visible in their images and the photometry does not appear
robust (e.g. $r$=24.2 mag, with $g-r$=--1.6) -- this may be the
counterpart but should be confirmed with deeper imaging.  The shorter
proposed wing is east of northern lobe. 

\smallskip\noindent{\bf (43) J1135--0737:} A faint optical counterpart to
this clear X-shaped radio source (main lobes are north-south) was recorded
in the USNO catalogue. A more prominent optical source ($b_J$=17.88 mag)
can be seen in the field overlapping with the outer edge of the southwest
wing. 

\smallskip\noindent{\bf (44) J1140+1057:} The SDSS identifies this as a
nearby ($z$=0.081) radio galaxy. The wing to the south of the eastern lobe
is more prominent than the one north of the western lobe. A bright optical
field source is seen off the edge of the eastern lobe. 

\smallskip\noindent{\bf (45) J1200+6105:} A wing to the south is clear but
its northern counterpart is not apparent in this double-lobed
(approximately east-west) radio source. The optical counterpart is near
the center of the radio source as expected. 

\smallskip\noindent{\bf (46) J1201--0703:} The radio source consists of a
northern pointing lobe (close to the optical counterpart) and an extensive
southeastern lobe.  Extended emission, perhaps from a wing, is seen to
stretch from the northeastern part of the northern lobe.  Possible wing
emission to the south seems to be associated with a $b$=19.5 mag source
found in the APM catalog [J2000: R.A.=12h01m26.93s, Dec.=--07d03m44.4s]
with a radio peak of 13 mJy/bm. 

\smallskip\noindent{\bf (47) J1202+4915:} There is no clear optical
counterpart in the DSS plates to this faint radio galaxy but one was found
($r$=21 mag) in the SDSS database. The FIRST image suggests the appearance
of a pair of faint east-west wings. 

\smallskip\noindent{\bf (48) J1206+3812:} This radio source was identified as
an $r$=18 mag quasar at high redshift ($z$=0.838) on the basis of MgII and
OII lines by \citet{vig90}. This is confirmed by the SDSS. If the X-shaped
morphology can be confirmed, this is by far the highest redshift X-shaped
source known. It shows a bright double-lobed radio morphology which is
quite symmetric. 

\smallskip\noindent{\bf (49) J1207+3352:} This B2 radio galaxy has been 
studied by \citet{par86} with the VLA and \citet{cap00} with HST.  It is
identified with a $z$=0.0788 galaxy in \citet{fan87}.  The optical and
radio core emissions peak very closely. The FIRST image shows similar
structure to what is seen in \citet{par86}. 

\smallskip\noindent{\bf (50) J1210--0341:} The radio lobes run in the NW-SE
direction with orthogonal low surface brightness wings quite apparent even
in the limited resolution FIRST image. The optical source near center is
identified with a $z$=0.26 galaxy \citep{mac99}.

\smallskip\noindent{\bf (51) J1210+1121:} The $z$=0.2 (SDSS) galaxy is
just offset from the southwest peak of the radio source giving this a very
lop-sided appearance. The proposed western wing is very extended at over 1
arcmin in length. 

\smallskip\noindent{\bf (52) J1211+4539:} A faint optical counterpart ($r$=22
mag) found in the SDSS database. This is too faint to be readily visible
in the DSS plates. There is a hint of wings for this symmetric radio double
(NW-SE active lobes), although they not very extended in the FIRST image. 

\smallskip\noindent{\bf (53) J1218+1955:} Symmetric radio source (NE-SW) with a
faint optical counterpart near its center. The wings are short (projected)
relative to the active lobes. 

\smallskip\noindent{\bf (54) J1227--0742:} The nearest plausible optical
counterpart is the nearby \citep[$z$=0.029;][]{fai92} $b$=15.0 mag galaxy
MCG -01-32-011 about 0.2' offset from the center of the radio source;
however, if this is the case, this would make the radio source quite
asymmetric and unusual looking. We rather favor that the optical
counterpart could be near the center of the radio source, in between the
two active lobes (NW-SE), and is hidden by the galaxy light from MCG
-01-32-011. In this case, the two extended diffuse east-west emission
would be interpreted as wings. 

\smallskip\noindent{\bf (55) J1227+2155:} A faint optical source near the
center of the radio source was found in the USNO and SDSS catalogues.
There appears to be two active lobes just off the east-west axis, with
more extended lower surface brightness wings of radio emission north and
south of the optical source. 

\smallskip\noindent{\bf (56) J1228+2642:} A pair of optical sources appear
overlapping the radio structure; we associate the radio source with the
brighter of the two (more northern) as it is more centrally located
relative to the radio emission. We imagine that the radio lobes run
north-south; the more extended western wing would be associated with the
southern lobe. 

\smallskip\noindent{\bf (57) J1232--0717:} A faint optical counterpart was
identified in the USNO catalog. The radio source is double-lobed in the
east-west direction with the more extended proposed wing to the south. 

\smallskip\noindent{\bf (58) J1247+4646:} A faint optical counterpart (SDSS
$r$=22 mag) to this radio source.  There are hints of wings to the NE and
SW. 

\smallskip\noindent{\bf (59) J1253+3435:} The estimated redshift of 0.034
\citep{bra05} may be underestimated -- it implies a lower radio luminosity
than would be expected from its FR-II morphology ($L_{\rm 1.4
GHz}$ = 10$^{24}$ W/Hz). The radio source has been mapped at higher
resolution than the FIRST image presented here by \citet{mac83}. Their map
shows the a more compact hot spot in the northeast lobe but failed to
detect the diffuse emission shown in the FIRST map. 

\smallskip\noindent{\bf (60) J1258+3227:} The $r$=17 mag (SDSS) source
close to the northwest lobe appears to be the parent object of this radio
galaxy. The wing to the south of this lobe is obvious but no wing is
apparent on the opposite lobe. 

\smallskip\noindent{\bf (61) J1309--0012:} The optical counterpart is
centered between the east-west radio lobes. It is much more apparent in
the R-band image of \citet{bes99}. They identify this as a radio galaxy at
$z$=0.42.  The wing to the south of the western lobe is seen also in their
5 GHz image of similar resolution to the FIRST image; there is a
suggestion of a short wing north of the eastern lobe in the FIRST image. 
The optical sources just overlapping with the eastern radio lobe is
suggested to be part of a cluster, SDSS CE J197.463455-00.213781 at an
estimated $z$=0.367 \citep{got02}. 

\smallskip\noindent{\bf (62) J1310+5458:} The central radio component in the
low resolution FIRST image is peaked (23 mJy/beam) very near the optical
counterpart. The active lobes form an east-west axis with the core
component, and the wing-like extensions are orthogonal to this. 

\smallskip\noindent{\bf (63) J1316+2427:} A faint optical counterpart is 
found in the USNO and SDSS catalogs at the center of this faint double
(approximately east-west) radio source. One can imagine a pair of short
faint wings to the north and south. 

\smallskip\noindent{\bf (64) J1327--0203:} The radio source is identified
with an SDSS galaxy at $z$=0.18. The north-south wings are quite prominent
although need higher resolution observations to improve the detailing of the
overall morphology. 

\smallskip\noindent{\bf (65) J1330--0206:} We associate the radio source with
the brightest ($g$=16.6; $z$=0.09 from SDSS) member of the southern
sub-cluster of Abell~1750 \citep{bee91}. This particular galaxy is near
the center of the double-lobed (approximately east-west) structure. The
galaxy at the southern edge of the eastern lobe is a fainter ($g$=19.2
mag) cluster member. 

\smallskip\noindent{\bf (66) J1339--0016:} \citet{dow86} pointed out that the
$z$=0.145 (SDSS) galaxy near the southern peak of this powerful extended
radio source (a.k.a. PKS~1137--000) is probably the optical counterpart,
rather than the $z$=1.818 quasar \citep{cro01} that is the faint optical
smudge at the eastern edge of a northern lobe component [J2000:
R.A.=13h39m33.9s, Dec.=--00d16m13s].  The optical galaxy is in between two
radio peaks toward the southern edge of the radio source. The galaxy is in
the field of the SDSS galaxy cluster SDSS CE J204.901352-00.280581 at and
estimated $z$=0.16 \citep{got02}. The western wing associated with the
northern lobe is quite prominent while the eastern wing of the southern
component takes a dramatic southerly bend after extending to the
east/northeast. 

\smallskip\noindent{\bf (67) J1342+2547:} The optical counterpart of this
roughly north-south radio source is closer to the fainter southern lobe.
The wings are quite clearly extended in roughly the east-west directions. 

\smallskip\noindent{\bf (68) J1345+5233:} We have found a faint smudge in the
DSS red plate very near the peak of this faint centrally-peaked radio
source. This source is catalogued in the SDSS database. In the low
resolution FIRST image, there appears to be four radio extensions directed
outward from the optical source. The east-west axis is more extended. 

\smallskip\noindent{\bf (69) J1348+4411:} The radio lobes run north-south and
are not quite symmetric with respect to the optical counterpart. The wings
are pretty clear -- the western wing off the southern lobe runs out to
$\sim$30\arcsec\ marked by the single contour (Figure~\ref{fig-new}). 

\smallskip\noindent{\bf (70) J1351+5559:} This is a distorted radio source 
with several galaxies in the near field determined to be at z$\simeq$0.07.
However, there is no redshift determination for the bright galaxy we have
associated with the central radio source. One can imagine the radio lobes
oriented roughly east-west and that there are wings to the south-north
with an extended spur of emission in the southern direction.  The two
neighboring galaxies to the north are probably physically associated: SDSS
J135142.56+555957.7 ($g$=17.8; $z$=0.06786$\pm$0.00008) and the slightly
brighter more northerly SDSS J135142.13+560004.5 ($g$=17.3;
$z$=0.07036$\pm$0.00019).  Another physically associated galaxy is about
2' further north (not shown), SDSS J135140.26+560125.1 ($g$=18.0;
$z$=0.06789$\pm$0.00015). 

\smallskip\noindent{\bf (71) J1353+0724:} There is a distracting gradient
across the DSS red image; otherwise, the optical counterpart is clearly
centered in between the radio lobes. A wing appears to the north,
associated with the fainter western lobe. 

\smallskip\noindent{\bf (72) J1406--0154:} There is a faint optical smudge in
the DSS red plate (SDSS $r$=21 mag) near where the two radio lobes cross.
In the FIRST image, there is emission suggestive of a wing to the south of
the western lobe and a shorter one north of the eastern lobe. 

\smallskip\noindent{\bf (73) J1406+0657:} There are clear extended wings
in the NW-SE direction roughly orthogonal to the axis of a double-lobed
radio structure. The central radio peak is coincident with an optical
source in the USNO and SDSS catalogues. 

\smallskip\noindent{\bf (74) J1408+0225:} The radio structure is barely
resolved in the FIRST image but is suggestive of having lobes running
east-west and that there is a more prominent wing to the south. The
optical counterpart, found in the USNO and SDSS catalogs, is coincident
with the FIRST radio image peak of 40.6 mJy/beam. 

\smallskip\noindent{\bf (75) J1411+0907:} A faint optical source in the USNO
and SDSS catalogues appears in between the two radio lobes. An extended
wing to the southwest of the southern radio lobe is obvious but one
associated with the northern lobe is not apparent in the FIRST image. 

\smallskip\noindent{\bf (76) J1424+2637:} This nearby B2 radio galaxy is
identified with MCG +05-34-033 \citep[$z$=0.037;][]{mil02} and has been
extensively studied. It has been imaged in the radio \citep[e.g.,
][]{der86}, with HST \citep{cap00}, and is an X-ray source \citep{can99}.
The wing north of the eastern lobe is shorter. 

\smallskip\noindent{\bf (77) J1430+5217:} The optical counterpart is
coincident with the central peak (26 mJy/beam) in the FIRST image. This
object is identified as a quasar at $z$=0.367 in the SDSS spectrum. The
southwest wing is slightly more extended than the northeastern one. 

\smallskip\noindent{\bf (78) J1433+0037:} A faint optical smudge in the DSS
plate and found in the SDSS catalog is coincident with a radio peak (4
mJy/beam) near the center of this radio galaxy. A wing is quite clear to
the east of the northern lobe, however, the ``wing'' from the southern lobe
runs to the south, not west as would have been expected. 

\smallskip\noindent{\bf (79) J1434+5906:} We find a faint smudge in the
DSS red plate centered between the two lobes oriented in the NE-SW
direction. The southeast wing is much more prominent.  In the field, we
see SDSS J143406.28+590704.5, a $g$=15.4 mag galaxy at $z$=0.040 (a faint
radio counterpart peaked at 0.6 mJy/beam in the FIRST image), prominently
30\arcsec\ away north of east of the radio galaxy. 

\smallskip\noindent{\bf (80) J1437+0834:} A faint optical counterpart was
catalogued by the USNO and SDSS near the center of double (east-west)
radio source. The more prominent wing is north of the western lobe. 

\smallskip\noindent{\bf (81) J1444+4147:} The optical counterpart is a
$z$=0.188 galaxy from the SDSS and is centered on the radio source. There
is a (SDSS $z$=0.255) galaxy in the field only 0.4 arcmin away overlapping
the eastern lobe.  We suggest the low surface brightness emissions to the 
north and south are wings.

\smallskip\noindent{\bf (82) J1454+2732:} There is no obvious optical
counterpart to this (north-south) radio source in the DSS image shown in
Figure~\ref{fig-new}, but a $r$=20 mag object is catalogued in the SDSS. 
The southern lobe is extended and the wing east of the northern lobe is
apparent. 

\smallskip\noindent{\bf (83) J1455+3237:} The optical field is a bit busy
with several field sources appearing in the DSS image. The optical source
we have chosen as the counterpart is a bright ($r$=16 mag) object at a
$z$=0.084 (SDSS) and is closest to the center of the radio source. The
main axis is roughly east-west and there is an extension to the north that
is suggestive of winged emission. 

\smallskip\noindent{\bf (84) J1456+2542:} A very faint $r$=20.6 mag
counterpart in the SDSS catalog is centered on the faint radio source. We
imagine the main lobes to be oriented north-south with the more prominent
wing east of the northern lobe. 

\smallskip\noindent{\bf (85) J1459+2903:} The mainly north-south radio source
shows a hint of additional extended emission east of the north lobe rather
than a sharp edge characteristic of X-shaped sources. The proposed wings
are east and west of the galaxy. This was previously studied with the VLA
and HST by Fanti and collaborators \citep[e.g.,][]{fan87,cap00} as part of
their studies of B2 radio galaxies and could very well be a normal radio
galaxy. The radio peak (22 mJy/beam) in the FIRST image is coincident with
the optical source identified with a $z$=0.146 galaxy \citep{gon00}. 

\smallskip\noindent{\bf (86) J1501+0752:} A faint smudge appears near the
center of the double-lobed radio source in the DSS plate.  The SDSS
catalog lists this smudge as a r$\sim$21 mag object and we have centered
the field on this source. There is a clear wing north of the northern lobe
but the wing associated with the southern lobe is not as obvious. 

\smallskip\noindent{\bf (87) J1515--0532:} The radio source is fairly
large in the sky ($\sim$1.5 arcmin in extent) and straddles a very faint
smudge in the DSS plates -- we have chosen to center the field on this
smudge catalogued in the APM.  The wings are short in projection (west of
the north lobe and east of south lobe) if they are indeed winged emission. 

\smallskip\noindent{\bf (88) J1522+4527:} There are two $z$=0.277 SDSS
galaxies overlapping the radio structure but neither are clearly the
nucleus of the radio source if this is a double-lobed radio galaxy (both
optical sources are near the outer edge of the radio lobes).  The more
centrally located (relative to the radio source) galaxy (SDSS
J152212.52+452804.7;  $g$=19.2) is only 0.2 arcmin to the north of the gap
between the radio double; the other is 0.9 arcmin away (SDSS
J152210.97+452708.7; $g$=20.4). There are hints of low surface-brightness
emission around the source, especially to the east, which may be wings. 

\smallskip\noindent{\bf (89) J1537+2648:} There is a clear optical
counterpart coincident with a central radio peak in between the two
(NW-SE) radio lobes. A very prominent wing associated with the northern
lobe stretches to the northeast but the one associated with the southern
one is not obvious in the FIRST image. 

\smallskip\noindent{\bf (90) J1600+2058:} This is a good example of an
X-shaped radio source with a morphology suggestive of a low-power FR-I
radio galaxy. The optical counterpart has not yet been spectroscopically
identified. 

\smallskip\noindent{\bf (91) J1603+5242:} The optical counterpart is aligned
with an obvious radio core in an edge-brightened (NW-SE) radio source. The
shorter wing is west of the southern lobe while the one associated with
the northern lobe is quite obvious. 

\smallskip\noindent{\bf (92) J1606+0000:} The radio source has been
identified with a nearby, $z$=0.059 galaxy \citep{bes99} whose ellipticity
is quite obvious in the DSS image. A high resolution 5 GHz VLA map shows
what appears to be a one-sided jet directed to east although there is a
mismatch between the brightest radio component and the center of the
optical galaxy in the overlay shown in \citet{bes99}.  The very extended
radio wings to the north and south revealed in the FIRST map do not appear
in their high resolution map. 

\smallskip\noindent{\bf (93) J1606+4517:} There is faint optical counterpart
($r$=20.5 mag) in the SDSS catalog centered on north-south double radio
source. The more prominent wing we propose is east of the northern lobe. 

\smallskip\noindent{\bf (94) J1614+2817:} The radio lobes are aligned
roughly east-west with a pair of prominent wings in the orthogonal
direction. The optical identification comes from \citet{mil01} who found
it to be nearby galaxy at $z$=0.107. 

\smallskip\noindent{\bf (95) J1625+2705:} This object is identified as a 
fairly distant ($z$=0.526) quasar in the SDSS and is a known X-ray source
\citep{wol01}. The radio morphology is centrally peaked in the FIRST image
with extended emission to the east/northeast suggestive of a wing. 

\smallskip\noindent{\bf (96) J1653+3115:} Two adjacent (NE-SW) very faint
optical smudges are seen in the DSS plate near the center of the radio
source -- the central radio peak (40 mJy/beam) is offset by
$\sim$2\arcsec\ from the closer of the two (the SW one). We imagine the
main active lobes running in the NW-SE direction. 

\smallskip\noindent{\bf (97) J1655+4551:} A faint uncatalogued optical smudge
is found in the DSS red plate toward the northeast part of the radio
source. We suggest the presence of at least one wing toward the west if
the main lobes are in the NW-SE direction. 

\smallskip\noindent{\bf (98) J1656+3952:} An optical counterpart was found in
the DSS image coincident with a central radio component (11.5 mJy/beam
peak in the FIRST image) between the roughly east-west lobes. One wing is
seen to the south of the eastern lobe but the wing in the opposite
direction is not as obvious in the FIRST image. 

\smallskip\noindent{\bf (99) J2226+0125:} The center of the radio source
(i.e. the gap between the two active, north-south lobes in the FIRST map)
is approximately midway ($\sim$7\arcsec) between two possible optical
counterparts.  Based on its position relative to the radio structure, the
northern $b_{\rm J}$=19.9 mag galaxy [J2000: R.A.=22h26m45.44s,
Dec.=+01d25m17.5s] is our best guess but this is not definitive. For
reference, the galaxy overlapping the southern lobe is a $b_{\rm J}$=19.5
mag galaxy [J2000:  R.A.=22h26m45.87s, Dec.=+01d25m05.8s], both from the
APM catalog.  Both are members of the Zwicky cluster ZwCl~2224.2+0109
\citep{zwi65}. Two wings are quite obvious with the one associated with
the southern lobe protruding more toward the outer edge of the lobe rather
than the side. 

\smallskip\noindent{\bf (100) J2359--1041:} We identify the $r$=19 mag
(SDSS) optical source near the central peak of the mainly east-west radio
source with the host. Two brighter optical field sources overlap with the
radio source: one just $\sim$5\arcsec\ north of the optical counterpart
and the other in the western lobe, 17\arcsec\ away.  The proposed wings
are north of the western lobe and south of the eastern lobe.

\section{Discussion\label{section-known}}

We address four issues pertinent to the search and study of X-shaped
radio sources. First, we gauge the success of our method by taking a
census of known examples from the literature to determine the efficiency
in which they are recovered by our method (\S~\ref{section-census}). Some
basic properties of the candidates are compared to those of the known
examples (\S~\ref{section-radiolum}). We then explore source orientation,
projection, and evolution effects on the observed morphology in
\S~\ref{section-appearance}, and briefly discuss the possible 
misidentifications in our search (\S~\ref{sec-misid}).

\subsection{Census and Success of our Method in Identifying Known 
Examples\label{section-census}}

Of the 19 known winged and X-shaped sources listed in Table~\ref{tbl-known}
(\S~\ref{section-def}), 11 have FIRST coverage in the current data release and
the efficiency in which they are recovered allow us to gauge the limitations of
our search process. The results can be summarized as follows (see
Table~\ref{tbl-known}, column 10): 
 
\begin{itemize}

\item Nine objects contain at least one component in the radio source
satisfying our selection criteria ($>$5 mJy/bm peak and FWHM$>$5\arcsec;
\S~\ref{section-search}).  Five are readily identifiable as X-shaped
sources in the FIRST maps \citep[4C+01.30 was actually originally
identified as an X-shaped source from its FIRST image;][]{wan03}. The
remaining four, the largest among the known examples, were ``missed''
because of their large angular sizes. Although their double lobed nature
are evident in some of the FIRST maps, their low surface brightness wing
emission were resolved out. 

\item In the two remaining cases, our peak flux and component size
thresholds are set too high and we would not have inspected their FIRST
field images.  The angular size of the double-lobed structure in J1357+4807
is small and none of its diffuse winged emission was distinguishable in the
FIRST map. In J2157+0037, its X-shaped nature is quite evident in the FIRST
map \citep[we identified it in a printed version in][]{zak04}.  However,
although its eastern lobe is very bright (50.8 mJy/bm peak), its deconvolved
size of 4.85\arcsec$\times$3.24\arcsec\ is just under our threshold of
5\arcsec. 

\end{itemize}

\noindent In one of the missed cases (4C+48.29), the FIRST image field
containing its southern lobe/hot spot complex was initially selected for
further scrutinization based on its double structure and surrounding
diffuse emission (Figure~\ref{fig-j1020}). It was subsequently removed
from our list because of its association with 4C+48.29 in the NVSS map. 

All 19 objects have NVSS coverage, and as realized in the 4C+48.29 case,
many of these are distinguishable as X-shaped sources in the
$\sim$45\arcsec\ resolution maps while they are too large to be
distinguished in the FIRST maps (see above). Conversely, many of the
smaller angular size ones identifiable in the FIRST images (e.g., 3C~63,
4C+01.30, 4C+04.40) are unresolved in the NVSS maps. This gives a
complementary approach of the two surveys for this work. However, short
wings in large angular size sources (e.g. 3C~192, 3C~379.1) are not
distinguishable in the low resolution NVSS maps, while the FIRST images
resolve out the diffuse emission (like in 3C~192). The main limiting
factor in any NVSS search is that it is sensitive only to bright,
prominently winged large angular size sources and the majority of these
have probably already been identified. See column~11 in
Table~\ref{tbl-known} for a summary of the FIRST and NVSS search results
of the known sources.

\subsection{Radio Luminosity\label{section-radiolum}}

\begin{figure}
\includegraphics[width=4.0in]{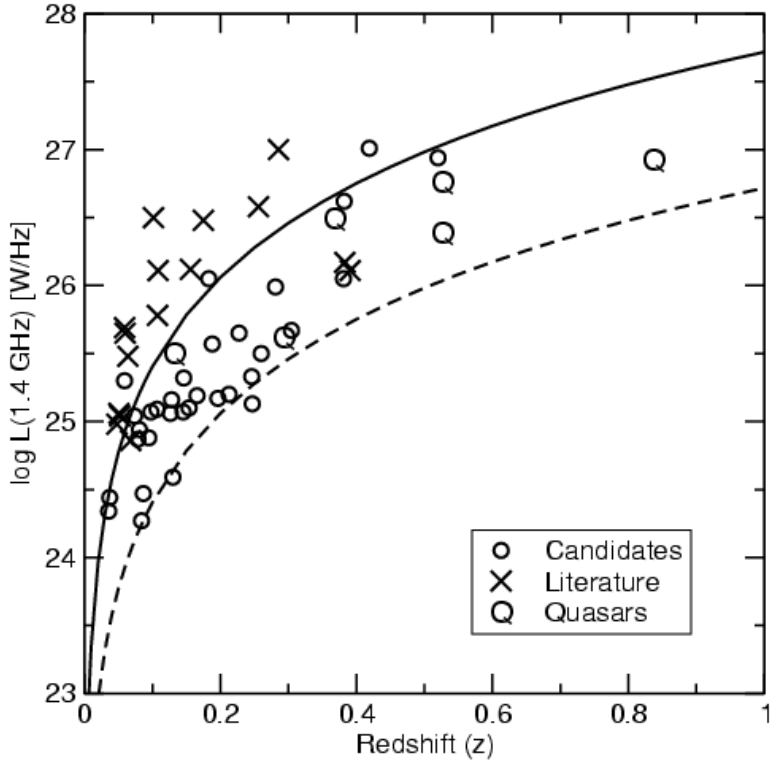}
\figcaption[f2.eps]{\label{fig-2}
The 1.4 GHz radio luminosity--redshift distribution of the
spectroscopically identified candidate (circles) and known (crosses)
X-shaped radio sources. Quasars are denoted with the letter Q (those
at z$>$0.3 were identified by us). The solid and dashed lines represent the
corresponding luminosity of a 1 Jy and 100 mJy radio source, respectively.}
\end{figure}

\begin{figure}
\includegraphics[width=6.5in]{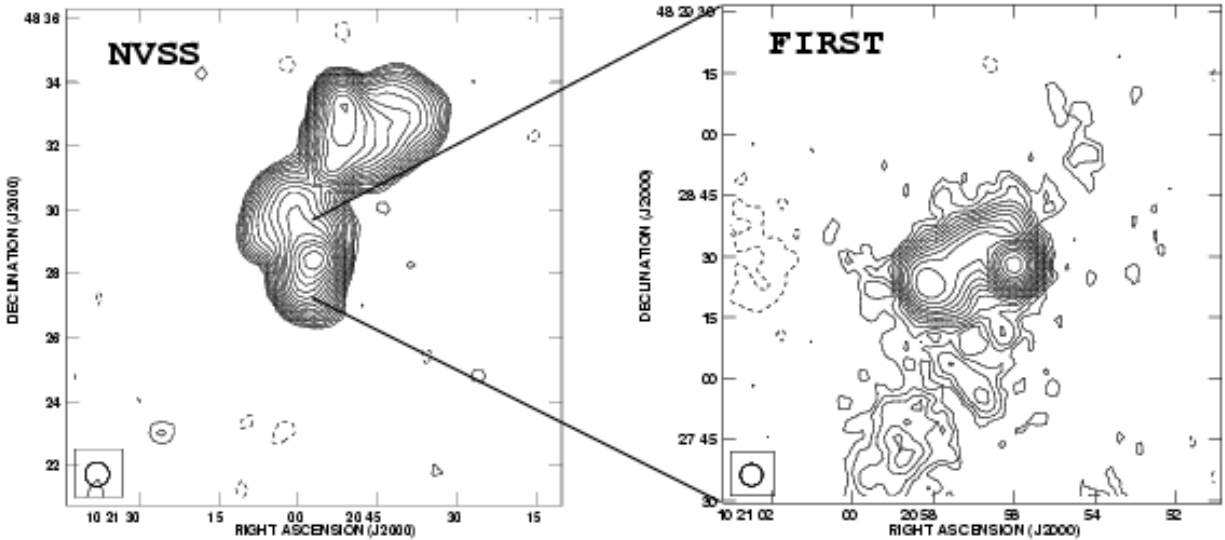}
\figcaption[fx.eps]{\label{fig-j1020}
Example of a FIRST image field [right panel] initially identified as an
X-shaped radio source candidate (imagined as east-west lobes and winged
emission to the south). A lower resolution map from NVSS [left panel]
revealed this to be a hot spot complex in the southern lobe of 4C+48.29,
coincidentally, a known X-shaped radio source (Table~\ref{tbl-known}). The
restoring beams of 45\arcsec\ (left) and 5.4\arcsec\ (right) are shown to the bottom
left of each panel.  The contour levels begin at 1.25 and 0.35 mJy/bm up
to peaks of 294.7 and 81.8 mJy/bm, respectively, increasing by
$\sqrt{2}$.} \end{figure}

\begin{figure}
\includegraphics[width=7.0in]{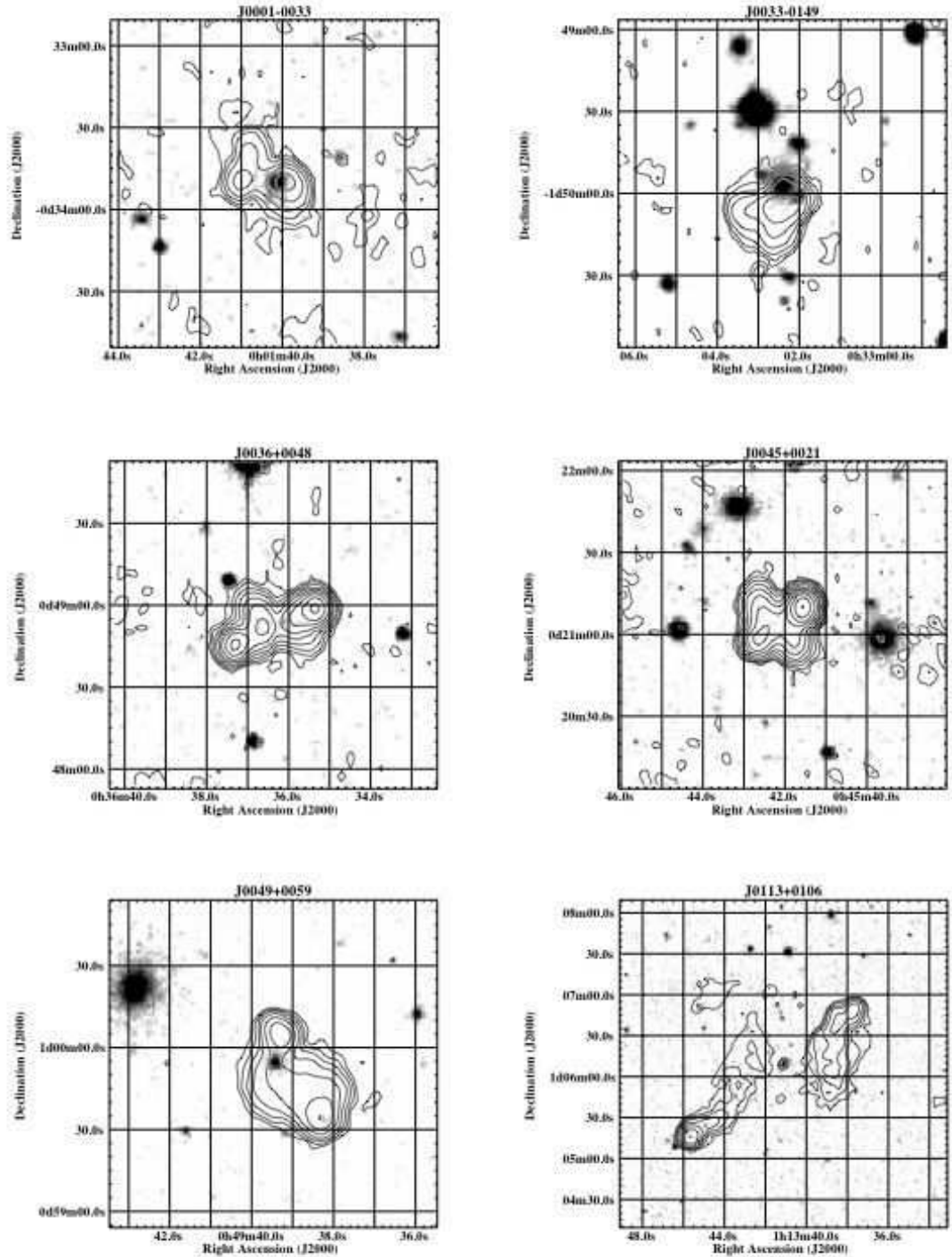}
\figcaption[f4.eps]{\label{fig-new}
VLA-FIRST 1.4 GHz images at 5.4\arcsec\ resolution (contours) of the 100  
candidate X-shaped radio sources overlaid with the DSS2 Red images 
(greyscale). The fields are centered on the optical counterparts when 
identified, otherwise, on positions based on the radio morphologies. The 
lowest radio contour plotted is 0.25 mJy/beam (about 2$\sigma$) increasing 
by factors of two. Image fields are 2$\times$2 arcmin$^{2}$ except for the 
two largest angular size sources, J0113+0106 and J1424+2637, where larger 
4$\times$4 arcmin$^{2}$ fields are shown.} \end{figure}

Figure~\ref{fig-2} shows the 1.4 GHz radio luminosity--redshift ($L-z$)
distribution of the spectroscopically identified (36) candidate winged and
X-shaped radio sources along with the (18) examples we compiled from
the literature; the quasars from both samples are marked as such.  We
see that the flux limit of our FIRST-based sample is effectively
$\sim$100 mJy, while the majority of the widely known examples
\citep[from the lists of][]{lea92,rot01,cap02} have 1.4 GHz flux
densities $\simgt$1 Jy. This means we are identifying X-shaped radio
sources with systematically lower radio luminosities for a given
redshift (by about 1 order of magnitude) than those from previous
catalogs.  In fact, the four most recently identified X-shaped sources
in the literature have flux densities comparable to our FIRST
candidates (J1130+0058 and J2157+0037 were actually identified in FIRST
radio maps). 

The lower flux threshold makes it possible to identify more distant, and
intrinsically lower luminosity (at a given redshift) radio sources as
X-shaped candidates. Simply by inspecting published maps, we extended the
redshift limit of known X-sources from $z\sim$0.3 to $\sim$0.4 with
J1357+4807 \citep{leh01} and J2157+0037 \citep{zak04}. We can extend our
census even further if we can confirm the morphologies in a number of our
candidates (four radio galaxies, J0115--0000, J0143--0118, J0941--0143,
and J1309--0012 at $z$=0.38 to 0.52, and four quasars from $z$=0.37 to
0.8). 

In such flux-limited samples, the lowest luminosity objects (like the
FR-I's) are naturally deselected out. This makes it important to confirm
if some of the less-luminous, more centrally peaked candidates (e.g. 
J0144--0830, J1345+5233, J1600+2058) do indeed have FR-I radio
morphologies \citep[like e.g. 3C~315,][]{ale87}.  Unfortunately, these are
not yet optically classified and are among the faintest and smaller
angular size radio sources in our compilation. Consequently, from
Figure~\ref{fig-2}, we see that to probe the high redshift counterparts of
the lower luminosity z$<$0.3 population ($\sim$10$^{24-26}$ W/Hz), we must
go fainter than $\simlt$100 mJy at 1.4 GHz. 

For the 16 X-shaped radio {\it galaxies} with spectroscopic identifications
comprising the ``literature'' sample (Table~\ref{tbl-known}), we find an
average 1.4 GHz luminosity of log~$L$ [W/Hz] = 25.85 (1$\sigma$ standard
deviation = 0.64; median=25.95). As X-shaped radio sources are known to have
radio luminosities close to the Fanaroff \& Riley type-I/II division, this
was expected\footnote{More precisely, they are found preferentially in
low-luminosity FR-II's \citep{lea92,den02}. The original \citet{fan74}
division of 6.3$\times$10$^{25}$ W/Hz/$h^{2}$ at 178 MHz translates to
$\sim$few $\times$10$^{25}$ W/Hz at 1.4 GHz for our adopted cosmology and
typical spectral index values around unity.}. Including the 32 candidates
identified radio-galaxies from our study brings the average luminosity even
closer to the FR-divide ($<$log~$L$$>$ [W/Hz] =25.49 with 1$\sigma$ =0.71;
median =25.31), although the large spread does not make this change
statistically significant. Including the quasars, where a portion of the
emission may be affected by Doppler beaming ($<$log~$L$$>$=26.3 W/Hz for the
six examples), has little affect on the values quoted as there are but a few
of these.  The connection to the Fanaroff-Riley luminosity division is
unclear although it is tempting to associate this with their transitory
appearances -- are these the long sought transition objects between FR-I's
and FR-II's? However, any definitive statements would be premature as for
instance, one is reminded that not all the candidates are yet confirmed
X-shaped radio sources. 

If one were however to pursue this association further, we must explain the
large range of observed luminosities of the X-shaped sources from
$\sim$10$^{24-27}$ W/Hz.  This may find a natural explanation in the fact
that the radio luminosity dividing the Fanaroff-Riley type-I from type-II
morphology sources is known to be dependent on the parent host galaxy
magnitude over a comparable radio luminosity range
\citep[e.g.,][]{led96}\footnote{Converting to our adopted cosmology, the
corresponding luminosities are $\sim$15--30$\%$ larger for the range of
redshifts considered.}. Further radio imaging of our candidates and a
dedicated host-galaxy imaging program can determine the proper placement of
X-shaped radio sources in the Owen-Ledlow plane, which may serve as a useful
diagnostic of the physical origin of these peculiar objects.

\subsection{Source Orientation, Projection and Evolution Effects on 
the Appearance of X-shaped Radio Sources \label{section-appearance}}

Observationally, one expects winged/X-shaped sources to be most apparent
when {\it both} the wings and active lobes are projected close to the
plane of the sky.  It is then no surprise that so far, the majority of
them have tended toward galaxies with narrow-line systems \citep{wan03}.
In unified schemes \citep{orr82,bar89}, these are the radio galaxies with
active lobes projected near the plane of the sky.  We must then consider
the effects of projection and source evolution on their appearance for any
attempt at a complete census of this class of objects.  This motivated us
to include candidates with shorter wings (i.e., wing to lobe length ratios
$<$0.8 like 3C~192 and 3C379.1; see \S~\ref{section-census}), more
amorphous looking ones, and sources with smaller angular sizes in
compiling our candidate list. 

Four of the spectroscopically identified X-shaped radio source candidates
are identified as (higher redshift; $z$=0.4--0.8) quasars, adding to the two
known examples (at lower-redshifts; Table~\ref{tbl-known}).  The relatively
few but increasing number of X-shaped quasars may be because their active
lobes are aligned closer to our line of sight and higher dynamic range is
required to detect the (unbeamed) low-surface brightness wings. 
Additionally, the number density of quasars peak at $z\sim$1--2
\citep[e.g.,][]{boy00}, so naively, we expect more X-shaped quasars to be
identified as we push to high-redshifts.  As an extreme case, \citet{mar06}
has recently pointed out that the $z$=2.065 quasar 0229+132 \citep{mur93}
may be one of the most extreme case of an aligned X-shaped radio source with
its large core-dominance. A significant population of X-shaped radio quasars
may have hitherto gone unnoticed. 

Source distance will also affect the appearance of an X-shaped radio
source.  For a given surface brightness (or constant wing luminosity), the
natural (1+z)$^4$ dimming of radiation makes it more difficult to detect
winged sources at high redshift. This may be why the two lower-redshift
($z$$<$0.3) X-shaped radio quasars from the literature \citep{wan03,lan06}
have more prominent wings than our higher redshift candidates.  In a
flux-limited sample, the higher-redshift sources tend toward having higher
luminosities (e.g., our quasar candidates are an order of magnitude more
luminous than the two lower-redshift ones from the literature); this and
cosmological dimming may explain the small number of high redshift
X-shaped radio galaxies and quasars identified so far (our high-$z$
candidates need improved radio imaging to confirm their X-shape). 

As alluded to earlier, the extent of the wings can simply be
foreshortened by projection.  In the active lobes, asymmetry/sidedness
of the opposite jets serve as a proxy of source orientation
\citep[e.g.,][]{orr82,bar89}. As an analogue, the pronounced asymmetry in
the relative prominence between the two wings in some objects may
indicate that the wings (in the realignment scenario, this is the
original source axis) are aligned close to our line of sight. 
Depolarization asymmetry in the wings could for example, then determine the
importance of projection in the wings -- this has been used as a proxy
of source orientation in the lobes of classical doubles
\citep{gar88,lai88}. 

If we can disentangle projection from intrinsic (i.e. source evolution)
effects, the extent of the wings may indicate the evolutionary stage of the
radio source. In a hydrodynamic origin, long wings means maximal back-flow
activity but short wings may either indicate the initial or terminal stages
of source activity. In the realignment scenario, the maximal extent and
surface brightness of the wings occurs shortly after realignment and
decreases thereafter. Radio spectral mapping of the lobes and wings
(spectral index and break frequency) can thus ``date'' the radio source
\citep{kle95,den02} to differentiate between these possibilities. 

In the evolution scenario, we should be able to see a continuum of wing to
lobe ratios with values so large as to create wing-dominated objects (at the
other extreme are lobe-dominated objects which are the classical doubles). 
These could masquerades as CSS or GPS sources \citep{ode98}. As an extreme
example, \citet{man04} recently found such a GPS source with a large scale
structure and a inner double source with flat radio spectra (they associate
these with a pair of AGN). Another possibility is that of the ``relaxed
doubles'' which have radio lobes appearing to be fading; two classic
examples (Hercules~A and 3C~310) have been found to have parsec-scale
structures misaligned with the large-scale lobes \citep{giz02}. Finally, the
borderline FR-I/II radio galaxy 3C~293 ($z$=0.0452) was identified early on
as a possible progenitor of the X-shaped sources \citep{bri81,den02}. Its
inner $\sim$2 kpc (0.88 kpc per arcsec) double-sided source straddles a flat
spectrum radio core which is coincident with the optical nucleus; this is
misaligned by $\sim$30\deg\ from the $\sim$200 kpc double larger structure.
\citet{den02} noted that 3C~293 shows the ``shortest ratio of active lobe
length to wing length is the only one which shows clear evidence for a
recent merger'' \citep[cf.][]{flo06}.

\subsection{Possible Misidentifications\label{sec-misid}}

With more sensitive, higher resolution radio imaging of these candidates, many
of them will turn out to be bona-fide X-shaped radio sources.  The remaining
will probably turn out to be simply winged sources, and a few are likely
unrelated to the X-shaped phenomenon. 

In particular, some of the objects may be ``bent-doubles,'' which have been
targeted previously in a similar search through the FIRST database
\citep{bla01}. Two (J0033--0149, J1111+4050) resemble narrow-angle tail (NAT)
radio sources \citep[e.g.,][]{ode85}. As other examples, the referee has cited
``J0813+4347, J1330-206 as plausible NATs and J0838+3253, J1339-0016 as possible
WATs'' (wide-angle tailed), and the author concurs. Finally, J0145--0159 and
J1040+5056 may turn out to be entirely normal FRI radio galaxies
(\S~\ref{sec-notes}), while J1043+3131 and J1351+5559 appear quite distorted in
the low resolution maps. Such sources will probably contaminate the final winged
and X-shaped source samples at the $\sim$10--20$\%$ level. This makes new radio
observations necessary to confirm or not the present morphological
identifications.

\section{Summary and Future Work\label{section-summary}}

A satisfactory explanation for the origin of the wings that give the
X-shaped radio galaxies their characteristic morphologies remains elusive.
This is mainly because of the small number studied in detail, and the
small total number of known examples available for systematic studies. 
This was the motivation for this FIRST-based search for new examples.
Candidates were selected by visual inspection of radio sources fields
containing resolved structure and imaged with sufficient fidelity to be
able to identify the winged structures. This paper presents an initial
sample of 100 candidates drawn from a subset ($\sim$1/5th of a total
$\sim$8,000 sources) of the currently data release matching our criteria. 

Of the 100 candidates, we found optical identifications for 94 of the radio
sources. Almost 40$\%$ of these have spectroscopic identifications available
from the literature, many from the SDSS database. In the radio
luminosity-redshift plane, it is clear that we are systematically probing an
order of magnitude lower radio luminosities than previous samples.  This allows
us to identify candidates at higher redshifts and a handful of objects with
morphologies characteristic of low-luminosity FR-I radio galaxies.  If
confirmed, such objects challenge current models for the formation/origin of
X-shaped radio sources -- powerful jet driven cocoons which are characteristic
of FRIIs are required to form the new active lobes in the merger scenario
\citep{gop03} and to drive the backflow in the hydrodynamic picture of
\citet{cap02}. 

We find an average radio luminosity of the combined sample from the
literature and our candidates to be close to the classical Fanaroff-Riley division,
albeit with a large spread. It may be a useful to project these
data onto the Owen-Ledlow (i.e.  host galaxy magnitude vs. radio luminosity)
plane to explore the relationship of these objects with the Fanaroff-Riley
division. 

This is only the first step in a systematic study of these objects. Higher
resolution imaging with the VLA are in progress and will detail the
candidates' morphologies to confirm or not their X-shaped nature. Follow-up
multi-frequency imaging of bona-fide X-shaped sources will be able to reveal
spectral differences between the wings and lobes and any spectral structure
to set upper limits on the particle ages in the wings, thus on the source
lifetime \citep[e.g.,][]{kle95,den02}. Observations can also detail the
varying prominence of the `Z' in the morphologies, which have been
taken as a sign of an ongoing merger \citep{gop03,zie05}. This scenario may 
lead to observable displacements in
their emission lines \citep{beg80} and between the radio core and optical
nucleus \citep{mad04} -- an optical spectroscopy program (in progress) and
sensitive VLBI imaging of the X-shaped radio sources with prominent radio
cores are necessary to test these interesting possibilities. 

An important issue in the formation of X-shaped radio sources is in their
environments. This can be constrained with X-ray data \citep{wor95,kra05}, an
examination of their optical fields (are these preferentially in groups or
clusters?), and radio polarization observations \citep{den02}. These can be
compared to the properties of classical radio galaxies \citep[e.g.,][]{zir97}.
Also, a determination of their host galaxy properties may provide additional
clues as to their origin \citep[e.g.,][]{ulr96,cap00}.  This work lays the
groundwork to produce an extensive sample of X-shaped sources in order to carry
out the follow-up work necessary to test the differing possibilities.

\acknowledgments
\noindent
\begin{center}
Acknowledgments
\end{center}

The inspiration to mine the FIRST database came after a colloquium talk by
David Helfand (in April 2005) at the MIT Kavli Institute (MKI), where the
author's fellowship was initially hosted. He is grateful to the MKI and
his current host, the KIPAC at Stanford for their hospitality.  Stephen
Healey and Hermine Landt provided valuable assistance and advice in the
optical identifications. Dan Harris, Christian Zier, and the anonymous referee 
gave very useful comments which improved the manuscript.



\begin{center}
Appendix: Notes on New X-shaped Radio Sources from the Literature
\end{center}

We provide brief descriptions of X-shaped radio sources not included in the
earlier compilations of \citet[][see \citet{mer02}]{lea92}, \citet{rot01}, and
\citet{cap02}. See Table~\ref{tbl-known} for the complete list.
Figure~\ref{fig-app} shows the radio/optical overlays for the 3/5 sources below
with the X-shaped structures apparent in the FIRST images. 

\smallskip\noindent{\bf J1130+0058:} This X-shaped radio source (4C+01.30)
was discovered in a FIRST map (Figure~\ref{fig-app}), and studied in
detail by \citet{wan03}. It is the first case of an X-source with a quasar
nucleus, at a moderate redshift of 0.132. 

\smallskip\noindent{\bf J1206+0406:} The X-shaped nature of J1206+0406 was
uncovered in a low-frequency (320 MHz) VLA map by \citet{jun00} and we
confirmed the morphology in the FIRST map (Figure~\ref{fig-app}). The
radio structure has an asymmetric appearance showing a one-sided jet to
the southwest and a diffuse lobe opposite of the core. The northwestern
wing is much more prominent than the one to the southeast.  No redshift is
available for J1206+0406 but we found a $r$=20.5 mag. optical counterpart
in the SDSS (J120620.12+040610.7, $g-r$=0.95). 

\smallskip\noindent{\bf J1357+4807:} We chanced upon a published 1.4 GHz
VLA map of J1357+4807 \citep[][Figure~2 therein]{leh01} which showed faint
wings on the NE-SW axis. They associate it with a $R$=19.7 mag object from
the APM; we confirm the identification as SDSS J135730.6+480741.7
($r$=20.1 mag.; $g-r$=0.9). We obtained a spectrum with the Hobby-Eberly
Telescope and found this to be a $z$=0.383 narrow-line radio galaxy (C.~C.
Cheung \& S.~E. Healey 2006, unpublished). 

\smallskip\noindent{\bf J2157+0037:} We also chanced upon a (FIRST) map of
this candidate type-II AGN in \citet{zak04} which shows the X-shape very
clearly (Figure~\ref{fig-app}). The optical counterpart (SDSS
J215731.43+003757) is a $r$=19.1 mag galaxy at $z$=0.391. 

\smallskip\noindent{\bf J2347+0852:} The X-shaped nature of J2347+0852 was
brought to our attention by H. Landt (2006, private communication) and the
VLA map is published in \citet{lan06}. It is associated with a $z$=0.292
quasar \citep{per98}.


{}

\clearpage

\pagestyle{empty}

\begin{sidewaystable}
\begin{footnotesize}
\caption[]{List of Known X-shaped* Radio Sources\label{tbl-known}}
\begin{center}
\begin{tabular}{lllccccccccl}
\hline \hline
J2000 Name & Other & $z$ & Ref. & $F_{\rm 1.4}$ & Radio maps & Leahy & Rottmann & Capetti &
FIRST ID? & NVSS ID? & Notes\\
(1)&(2)&(3)&(4)&(5)&(6)&(7)&(8)&(9)&(10)&(11)&(12)\\
\hline
J0009+1244 & 4C+12.03     	  &0.156* & P84 &  1982 & L91,L06         &\ch & \ch & \ch &..   &Yes &           \\      
J0058+2651 & NGC326, B2~0055+26   &0.0477 & W99 &  1782 & E78,M01         &\ch & \ch & ..  &..   &Yes &Double Galaxy \\
J0148+5332 & 3C52         	  &0.2854 & S85 &  3910 & L84,L86,A87,L06 &\ch & \ch & \ch &..   &No  &           \\
J0220--0156& 3C63       	  &0.175  & S80 &  3531 & B88,H98         &..  & ..  & \ch &Yes  &No  &           \\
J0516+2458 & 3C136.1	          &0.064  & S85 &  3073 & L84,L86,A87,L06 &\ch & \ch & \ch &..   &Yes &           \\
J0805+2409* & 3C192   	          &0.0598 & S66 &  5158 & B88,L97,D99,L06 &..  & ..  & \ch &r    &No  & Small wings  \\
J0831+3219 & 4C+32.25, B2~0828+32 &0.0507 & D86 &  1886 & P85,F87,M94,K95,L06&\ch & \ch & \ch &r & Yes&          \\
J0941+3944 & 3C223.1      	  &0.1075 & S66 &  2034 & B92,D02,L05     &\ch & \ch & \ch &Yes  &Yes &           \\
J1020+4831 & 4C+48.29     	  &0.052  & M79 &  1727 & V82,L06         &\ch & ..  & ..  &r    &Yes &           \\
J1101+1640 & Abell~1145, 1059+169 &0.0680 & O97 &   640 & O92,O97,L06     &\ch & ..  & ..  &r    &Yes &           \\
J1130+0058 & 4C+01.30             &0.1325 & W03 &   675 & W03             &..  & ..  & ..  &Yes  &No  & Quasar    \\
J1206+0406 & 4C+04.40, 1203+043   &..     & ..  &  1501 & M92,J00         &..  & ..  & ..  &Yes  &No  &           \\
J1357+4807 & ..                   &0.383  & C06 &   288 & L01             &..  & ..  & ..  &No   &No  &           \\
J1513+2607 & 3C315        	  &0.1083 & S65 &  4329 & L84,L86,A87,L06 &\ch & \ch & \ch &Yes  &Yes & Double Galaxy \\
J1824+7420* & 3C379.1              &0.2560 & S76 &  1879 & M85,Spa85       &..  & \ch & ..  &..   &No  & Small wings  \\
J1952+0230 & 3C403                &0.059  & S72 &  5900 & B92,D02,K05,L06 &\ch & \ch & \ch &..   &Yes &           \\
J2123+2504 & 3C433        	  &0.1016 & S65 & 11940 & V83,B92,L06     &\ch & ..  & ..  &..   &No  &           \\
J2157+0037 & ..			  &0.3907 & Z04 &   242 & Z04             &..  & ..  & ..  &p,s  &Yes &     \\
J2347+0852 & ..   	          &0.292  & P98 &   155 & Lan06           &..  & ..  & ..  &..   &No  & Quasar    \\
\hline \hline
\end{tabular}
\end{center}
The asterick (*) denotes the two sources with shorter winged 
structures than the other bona-fide X-shaped sources. As these 
were included in the known lists of X-sources, we list them here with the aim of summarizing these works.\\
(1, 2): Names based on J2000.0 coordinates and other common catalogue names.
(3, 4): Redshift ($z$) and original references. A new redshift was determined
for J1357+4807 (see references).
*Several works \citep[e.g.,][]{hec94,rot01,mer02} list this object as 
$z$=0.110, which was estimated from the object's apparent $r$-magnitude by 
\citet{lai83}. The quoted value here is derived from spectroscopic 
measurements of emission lines.\\
(5) Radio 1.4 GHz radio flux density (mJy). (6) Representative radio maps from the literature. See \citet{rot01}
for additional maps of sources from his list. \\
(7, 8, 9): Indicates objects appearing in the lists of \citet{lea92} (as 
given in \citet{mer02}), \citet{rot01}, and \citet{cap02}.\\
(10, 11): Indicates whether the X-shaped morphologies were distinguishable 
in the FIRST and NVSS maps.
In the FIRST column, the codes, p and s, meant it failed to 
meet one or both of our search criteria: p=peak below threshold of 5 
mJy/bm, but size large enough; 
s=size smaller than 5\arcsec\ threshold, but peak high enough. 
r= did meet both of our search criteria but the overall source 
angular size is too large and much of the diffuse emission was resolved 
out in the FIRST map. See \S~\ref{section-census} for a description. \\
\\
References: 
A87=\citet{ale87},
B88=\citet{bau88},
B92=\citet{bla92},
C06=C.C. Cheung \& S.E. Healey (2006, unpublished) -- see Appendix, 
D86=\citet{der86},
D99=\citet{den99},
D02=\citet{den02},
E78=\citet{eke78},
F87=\citet{fan87},
H98=\citet{har98},
J00=\citet{jun00},
K95=\citet{kle95},
K05=\citet{kra05},
L84=\citet{lea84},
L86=\citet{lea86},
L91=\citet{lea91},
L97=\citet{lea97},
L04=\citet{leh01},
L05=\citet{lal05},
L06=\citet{lal06},
Lan06=\citet{lan06},
M79=\citet{mil79},
M85=\citet{mye85},
M92=\citet{man92},
M94=\citet{mac94},
M01=\citet{mur01},      
O92=\citet{owe92},
O97=\citet{owe97},
P84=\citet{per84},
P85=\citet{par85},
P98=\citet{per98},
S65=\citet{sch65},
S66=\citet{san66},
S72=\citet{san72},
S76=\citet{smi76},
S80=\citet{smi80},
Spa85=\citet{spa85},
S85=\citet{spi85},
V82=\citet{van82},
V83=\citet{van83},
W99=\citet{wer99},
W03=\citet{wan03},
Z04=\citet{zak04}. 
\end{footnotesize}
\end{sidewaystable}

\clearpage

\begin{sidewaystable}
\begin{footnotesize}
\caption[]{Candidate X-shaped Radio Sources from FIRST\label{tbl-new}\\
{\bf Incomplete table on astro-ph. Go to http://www.stanford.edu/$\sim$teddy3c/Preprints/}
}
\begin{center}
\begin{tabular}{clllccccccrrrcccl}
\hline \hline
Catalog & Name & R.A. & Dec. & mag. & $g-r$ & Ref. & ID & $z$ & Ref. &
$F_{0.365}$ & $F_{1.4}$ & $F_{4.9}$ & Ref.$_{4.9}$ &   
$\alpha$$^{0.365}_{1.4}$ & $\alpha$$^{1.4}_{4.9}$ & Other Catalogs \\     
Number & & J2000.0 & J2000.0 & & & & & & & [mJy] & [mJy] & [mJy] & & & & \\
(1)&(2)&(3)&(4)&(5)&(6)&(7)&(8)&(9)&(10)&(11)&(12)&(13)&(14)&(15)&(16)&(17)\\
\hline
  1 & J0001--0033& 00 01 40.18 &--00 33 50.6  & 17.4   & 1.5  & SDSS &  G  &  0.2469 &  SDSS  &  ..   &    73 & 50  & PMN & ..   &  0.30 &                   \\  
  2 & J0033--0149& 00 33 02.41 &--01 49 56.6  & 14.1*  & ..   & USNO &  G  &  0.1301 &  6dF   &  169  &    87 & ..  & ..  & 0.49 &  ..   &                   \\
  3 & J0036+0048 & 00 36 36.21 & +00 48 53.4  & 20.4   & 1.2  & SDSS &  .. &  ..     &  ..    &  758  &   280 & 89  & PMN & 0.74 &  0.91 &                   \\
  4 & J0045+0021 & 00 45 42.11 & +00 21 05.5  & 21.5   & ..   & D89  &  .. &  ..     &  ..    & 1677  &   509 &159  & PMN & 0.89 &  0.93 & 4C-00.05, PKS     \\
  5 & J0049+0059 & 00 49 39.45 & +00 59 53.8  & 18.1   & 1.7  & SDSS &  G  &  0.3044 &  SDSS  &  427  &   155 & 77  & PMN & 0.75 &  0.56 &                   \\
    \\                                                                                                           
  6 & J0113+0106 & 01 13 41.11 & +01 06 08.5  & 18.1   & 1.3  & SDSS &  G  &  0.281  &  L00   &  ..   &   391 & 121 & PMN & ..   &  0.94 &                   \\
  7 & J0115--0000& 01 15 27.37 &--00 00 01.5  & 20.5   & 1.4  & SDSS &  G  &  0.381  &  L00   &  ..   &   222 & 54  & B91 & ..   &  1.13 & 4C-00.07, PKS     \\
  8 & J0143--0119& 01 43 16.75 &--01 19 00.8  & 19.3   & ..   & USNO &  G  &  0.520  &  L00   & 1978  &   823 & 340 & PMN & 0.65 &  0.71 & 4C-01.09, PKS     \\
  9 & J0144--0830& 01 44 09.98 &--08 30 02.8  & 18.6   & 1.2  & SDSS &  .. &  ..     &  ..    &  ..   &    47 & ..  & ..  & ..   &  ..   &                   \\
 10 & J0145--0159& 01 45 19.99 &--01 59 47.9  & 13.4*  & ..   & USNO &  G  &  0.1264 &  6dF   &  472  &   272 & 97  & PMN & 0.41 &  0.82 &                   \\
  \\                                                                                                           
 11 & J0147--0851& 01 47 19.28 &--08 51 19.6  & 20.2   & 1.1  & SDSS &  .. &  ..     &  ..    &  839  &   306 & 115 & PMN & 0.75 &  0.78 &                   \\
 12 & J0211--0920& 02 11 46.96 &--09 20 36.6  & 18.5   & 1.3  & SDSS &  .. &  ..     &  ..    &  377  &   180 & ..  & ..  & 0.55 &  ..   &                   \\
 13 & J0225--0738& 02 25 08.62 &--07 38 49.1  & 24.3   & 0.5  & SDSS &  .. &  ..     &  ..    &  916  &   318 & 131 & PMN & 0.79 &  0.71 & PKS               \\
 14 & J0702+5002 & 07 02 47.92 & +50 02 05.3  & 15.5*  & ..   & USNO &  G  &  0.0946 &  H05   &  761  &   334 & 115 & G91 & 0.61 &  0.85 & 6C                \\
 15 & J0725+5835 & 07 25 32.27 & +58 35 27.4  & 19.9   & ..   & USNO &  .. &  ..     &  ..    &  450  &   178 & 45  & G91 & 0.69 &  1.10 & 6C, 8C            \\
      \\                                                                                                           
 16 & J0805+4854 & 08 05 44.0  & +48 54 58    & ..     & ..   & FIRST&  .. &  ..     &  ..    &  ..   &    34 & ..  & ..  & ..   &  ..   &                   \\
 17 & J0813+4347 & 08 13 00.11 & +43 47 48.5  & 16.1   & 1.1  & SDSS &  G  &  0.1282 &  SDSS  &  624  &   333 & 156 & G91 & 0.47 &  0.61 & 6C, B3            \\
 18 & J0821+2922 & 08 21 49.60 & +29 22 44.4  & 20.2   & 0.9  & SDSS &  G  &  0.246  &  W03   &  478  &   117 & 27  & G91 & 1.05 &  1.17 & 5C07.178, B2      \\
 19 & J0836+3125 & 08 36 35.46 & +31 25 51.2  & 19.8   & 2.0  & SDSS &  .. &  ..     &  ..    &  524  &   291 & 111 & G91 & 0.44 &  0.77 & 6C                \\
 20 & J0838+3253 & 08 38 44.61 & +32 53 11.8  & 16.9   & 1.4  & SDSS &  G  &  0.2127 &  SDSS  &  ..   &   121 & 60  & G91 & ..   &  0.56 & 6C                \\
\hline \hline
\end{tabular}
\end{center}
(2, 3, 4): Name based on J2000.0 equinox positions in right ascension (R.A.) 
and declination (Dec.). Positions are of the optical counterparts when 
identified in the SDSS (most widely adopted), USNO, and APM catalogs, or 
other published works (see Column 7). When the optical counterparts were not 
catalogued, less precise positions were estimated directly from the DSS or 
FIRST images (Figure~\ref{fig-new}). \\
(5, 6, 7): Observed optical magnitudes (mag.) in the $r$-band. SDSS model 
magnitudes were preferred whenever available. 
USNO magnitudes are valid for point sources only so obviously extended 
objects in the DSS images are marked with an asterisk (*). SDSS $g-r$ colors 
are listed when available and anomalously large $g-r$ values are indicated 
(*). Inspection of SDSS data at their five-bands show inconsistent $g$-band 
magnitudes in these cases: more reliable colors are $u-r$=1.7 (J1247+4646) 
and 2.2 (J1411+0907). \\
(8, 9, 10): Spectroscopic identifications (ID; G=radio galaxy, Q=quasar), 
redshifts ($z$) and references. \\
(11, 12, 13, 14): Integrated radio flux density measurements at 365 MHz from 
the TXS survey, 1.4 GHz from NVSS, and at 4.9 GHz from the references listed in 
column (14): Green Bank (B91 and G91), PMN, or PKS catalogs. See 
\S~\ref{sec-radio} for further explanation.\\
(15, 16): Spectral index ($\alpha$) calculated between 365 MHz and 1.4 GHz 
(column 15) and between 1.4 and 4.9 GHz (column 16) where 
$F_{\nu}\propto\nu^{-\alpha}$.\\
(17): Other names and (mostly radio; the Cambridge, 4C, 5C, 6C, 7C, 8C, and, 
Bologna, B2 and B3) catalogs in which these objects appear.\\
\\
References: B99=\citet{bes99}, D89=\citet{dun89}, F87=\citet{fan87},
F99=\citet{fal99}, G00=\citet{gon00}, H05=S.E. Healey (2005, private
communication), L00=\citet{lac00}, M99=\citet{mac99}, M01=\citet{mil01},
M02=\citet{mil02}, O95=\citet{owe95}, S79=\citet{spi79}, V90=\citet{vig90},
W03=\citet{wil03}.\\
\\
Catalogs: 6dF=\citet{jon05}, APM=\citet{mad90}, SDSS=\citet{ade07},
USNO=\citet{mon03}, FIRST=\citet{bec95}, Green Bank: B91=\citet{bec91} and
G91=\citet{gre91}, NVSS=\citet{con98}, PMN=\citet{gri94,gri95},
PKS=\citet{wri90}, TXS=\citet{dou96}.
\end{footnotesize}
\end{sidewaystable}

\clearpage

\pagestyle{plaintop}


\clearpage

\clearpage

\addtocounter{figure}{0}
\begin{figure} \includegraphics[width=7.0in]{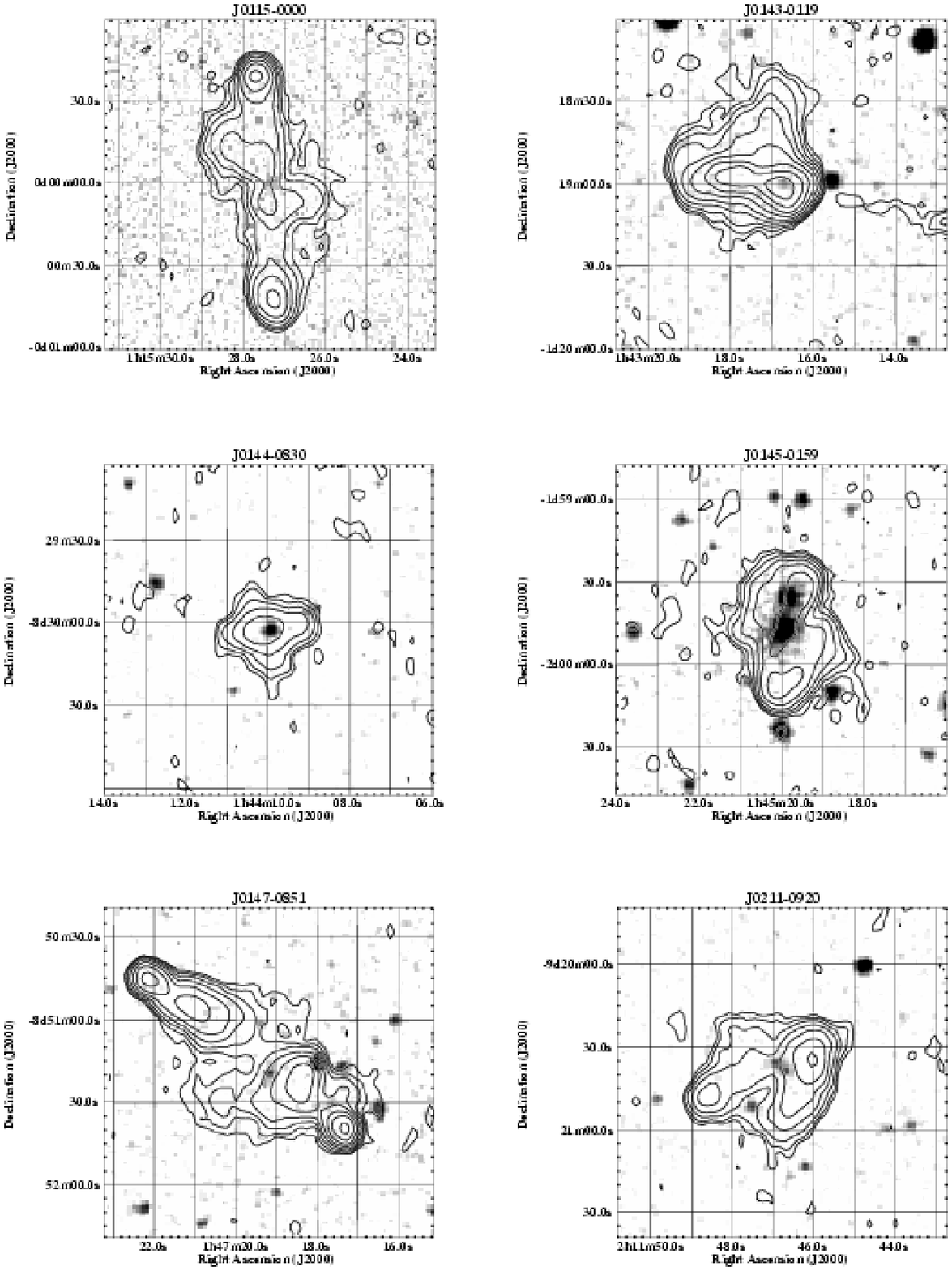} \end{figure}
\addtocounter{figure}{0}
\begin{figure} \includegraphics[width=7.0in]{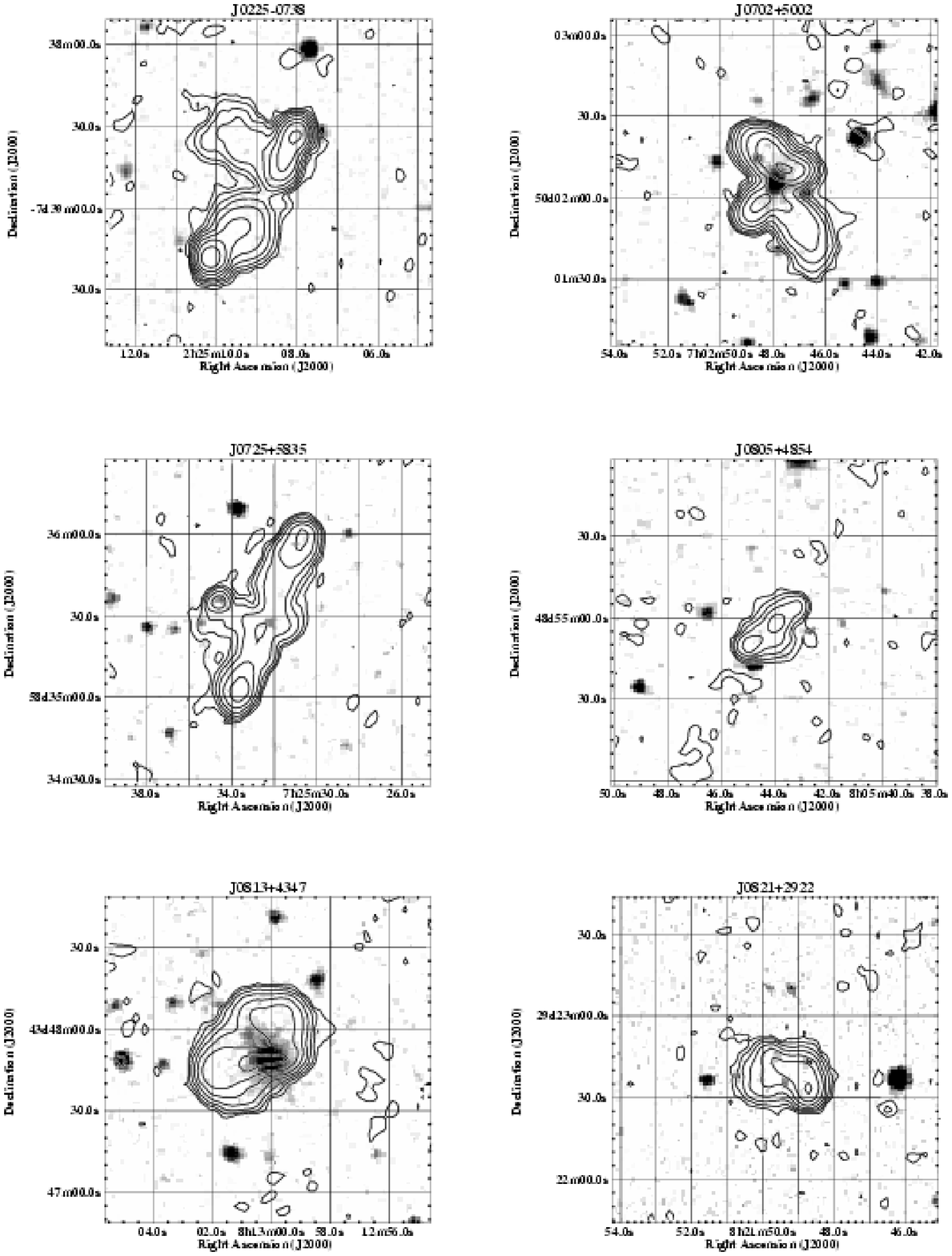} \end{figure}
\addtocounter{figure}{0}
\begin{figure} \includegraphics[width=7.0in]{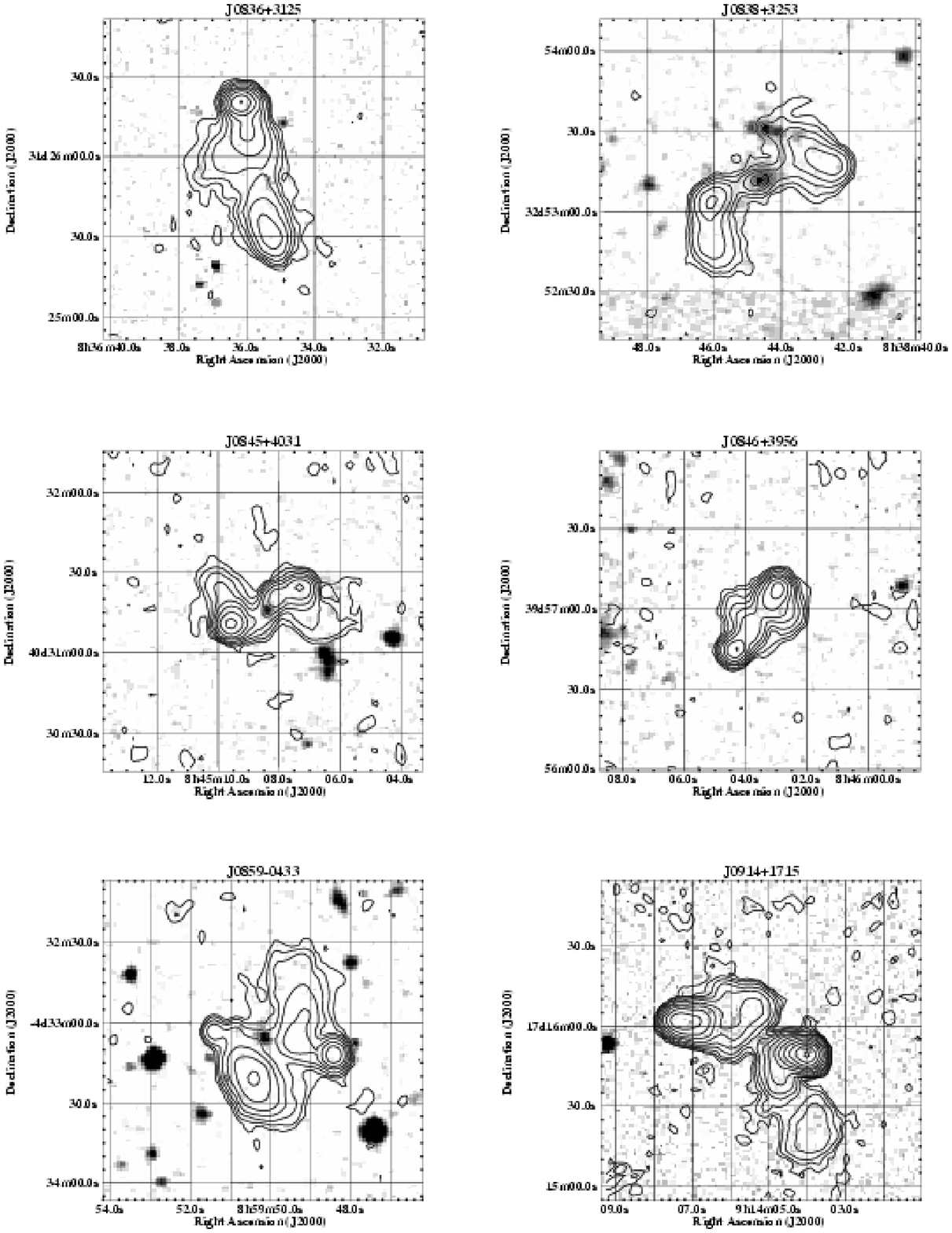} \end{figure}
\addtocounter{figure}{0}
\begin{figure} \includegraphics[width=7.0in]{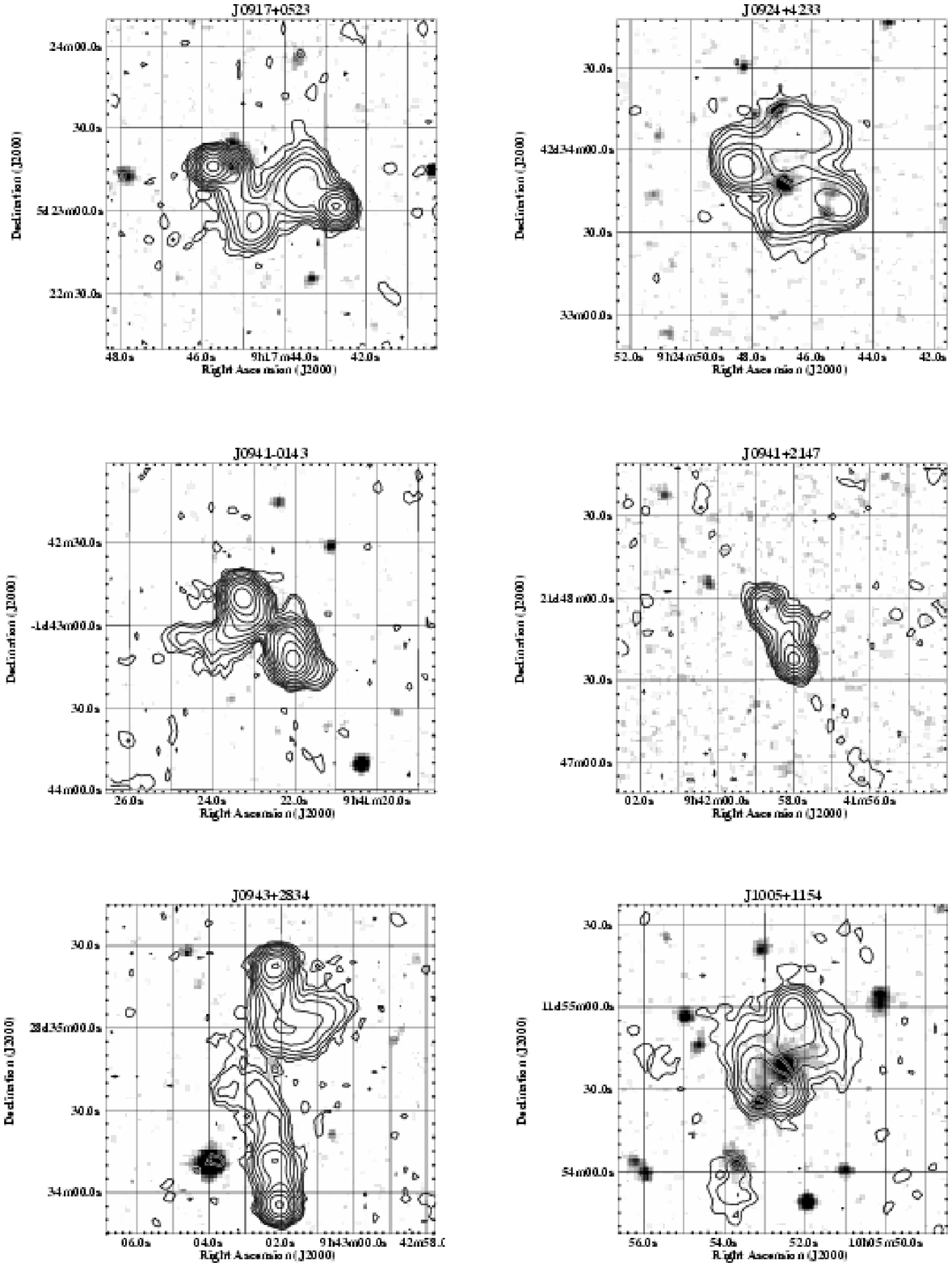} \end{figure}
\addtocounter{figure}{0}
\begin{figure} \includegraphics[width=7.0in]{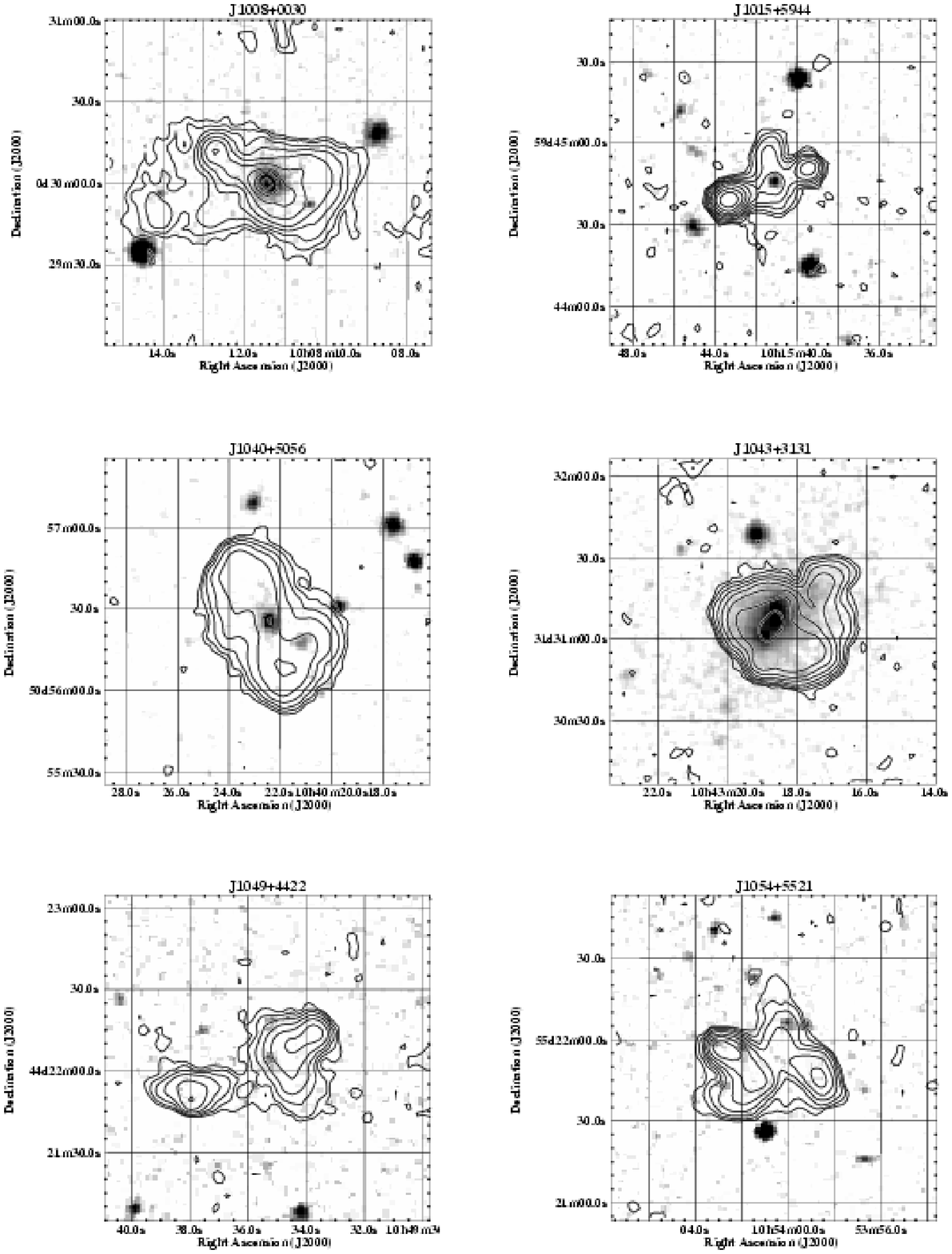} \end{figure}
\addtocounter{figure}{0}
\begin{figure} \includegraphics[width=7.0in]{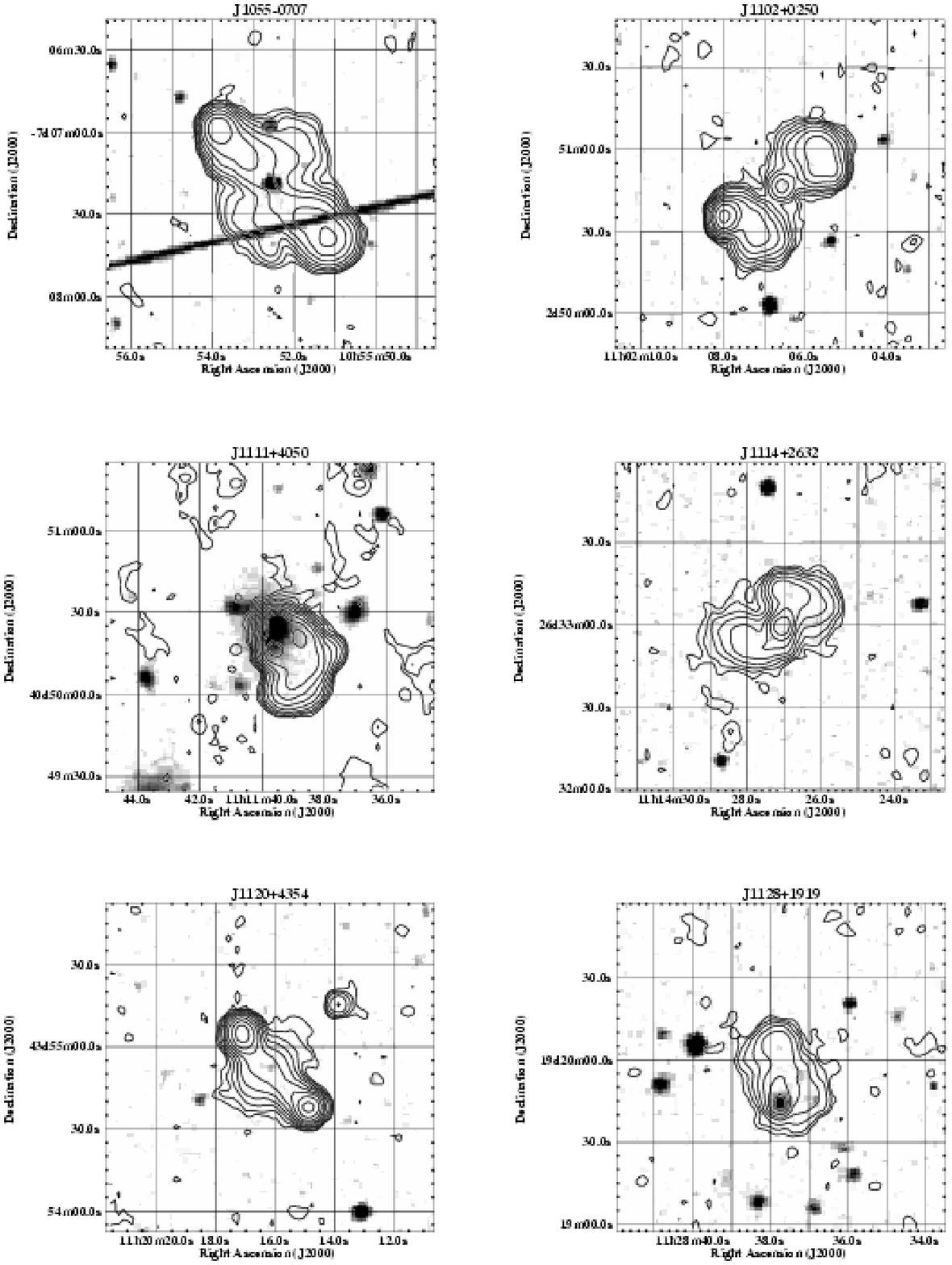} \end{figure}
\addtocounter{figure}{0}
\begin{figure} \includegraphics[width=7.0in]{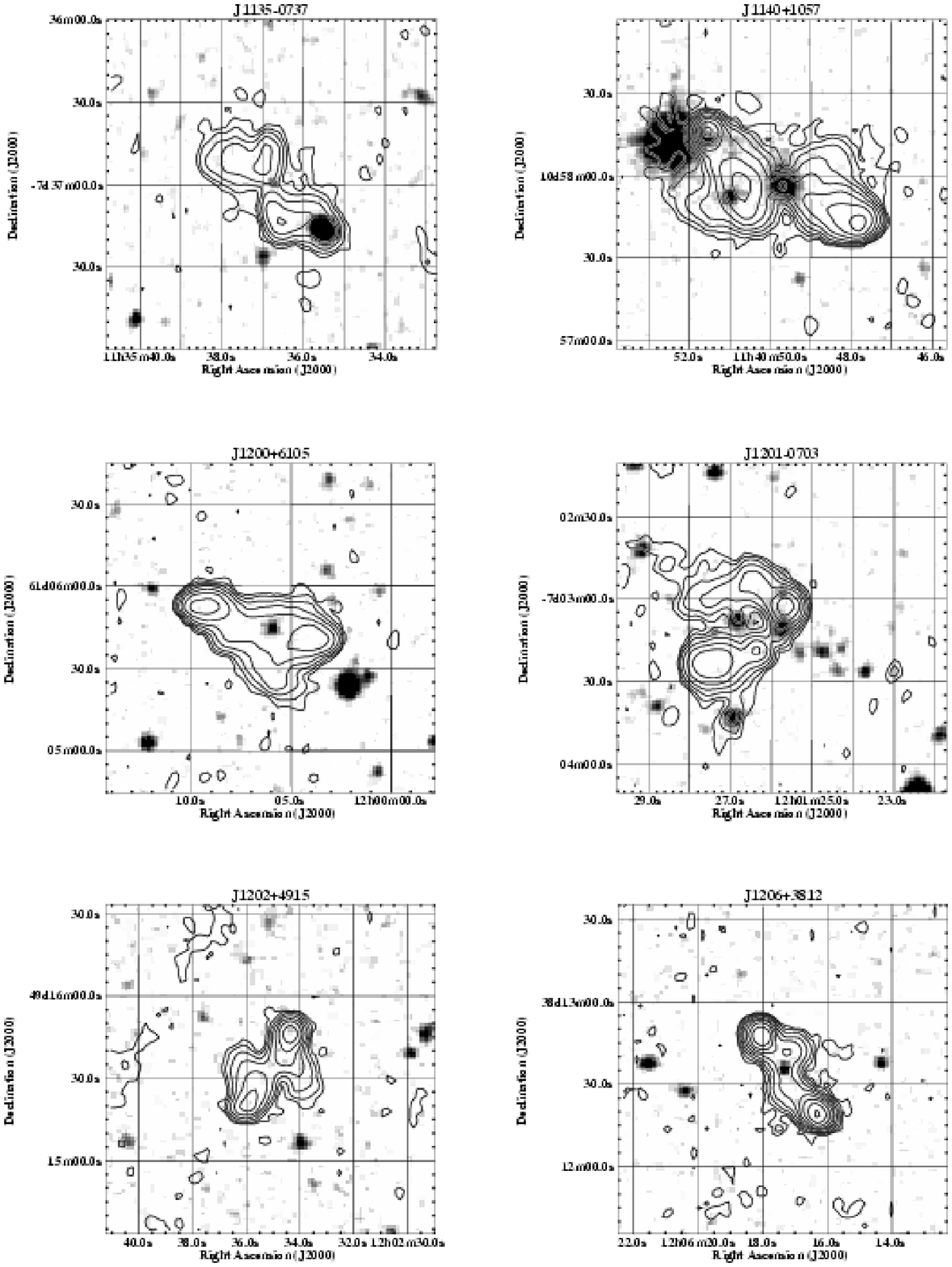} \end{figure}
\addtocounter{figure}{0}
\begin{figure} \includegraphics[width=7.0in]{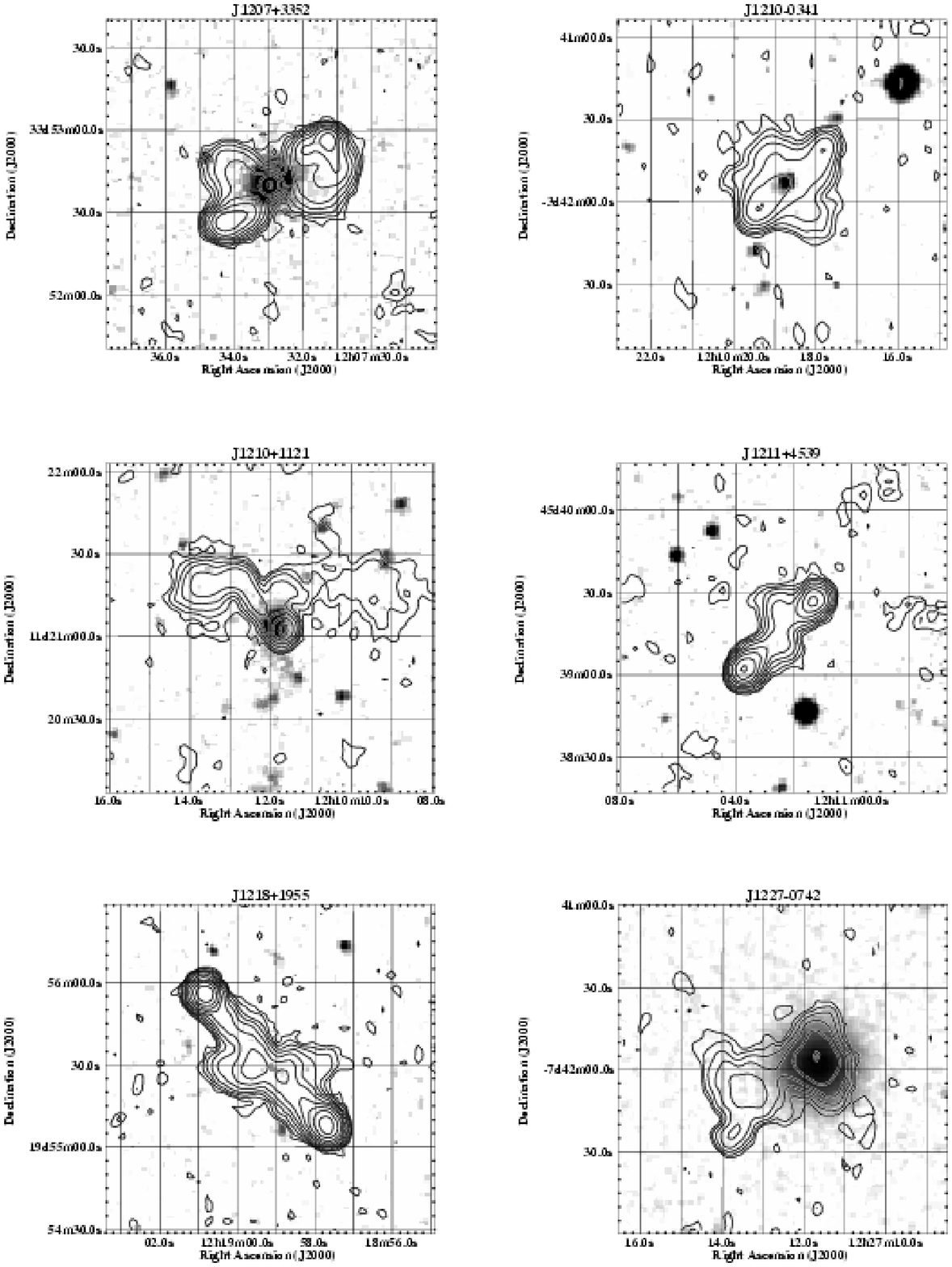} \end{figure}
\addtocounter{figure}{0}
\begin{figure} \includegraphics[width=7.0in]{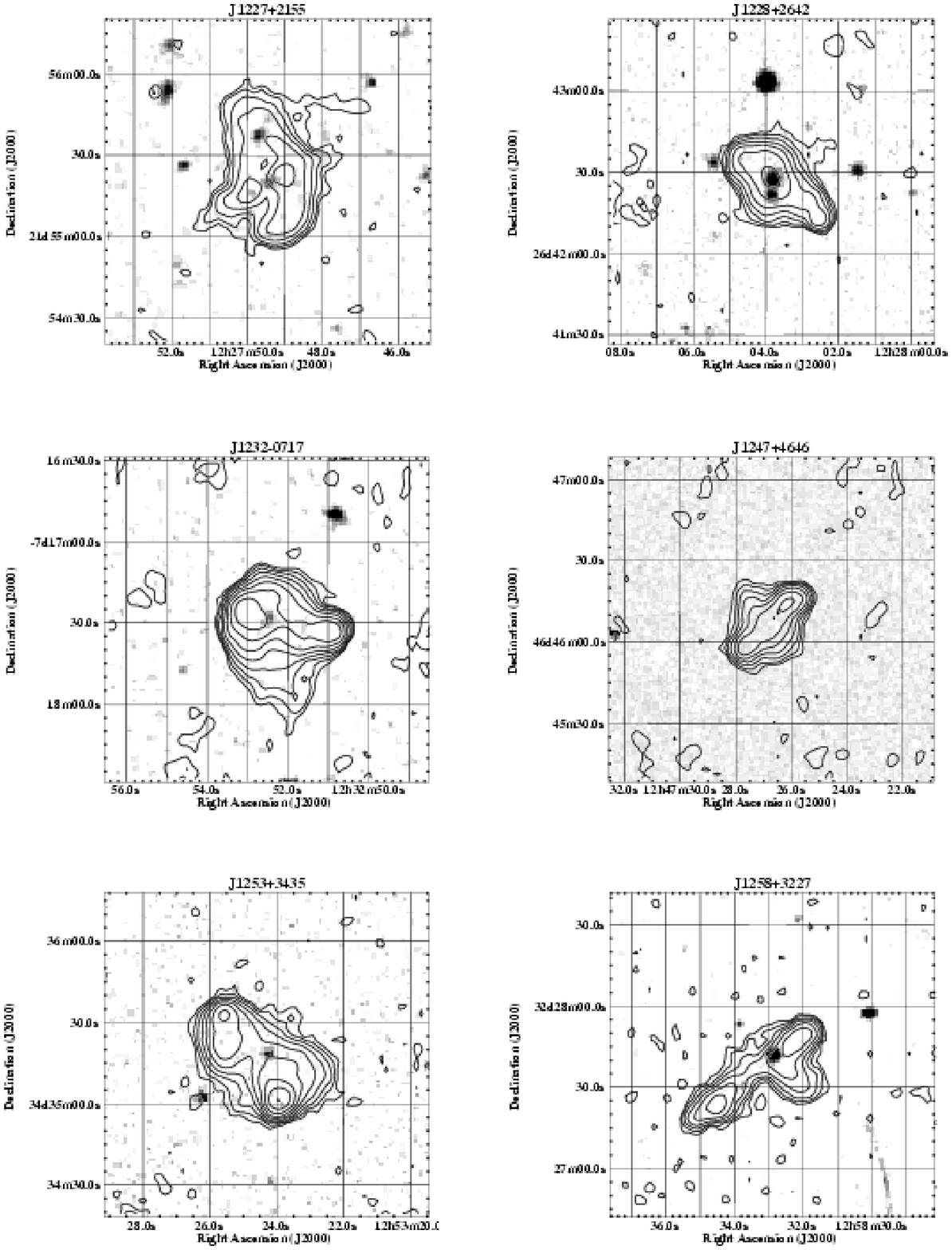} \end{figure}
\addtocounter{figure}{0}
\begin{figure} \includegraphics[width=7.0in]{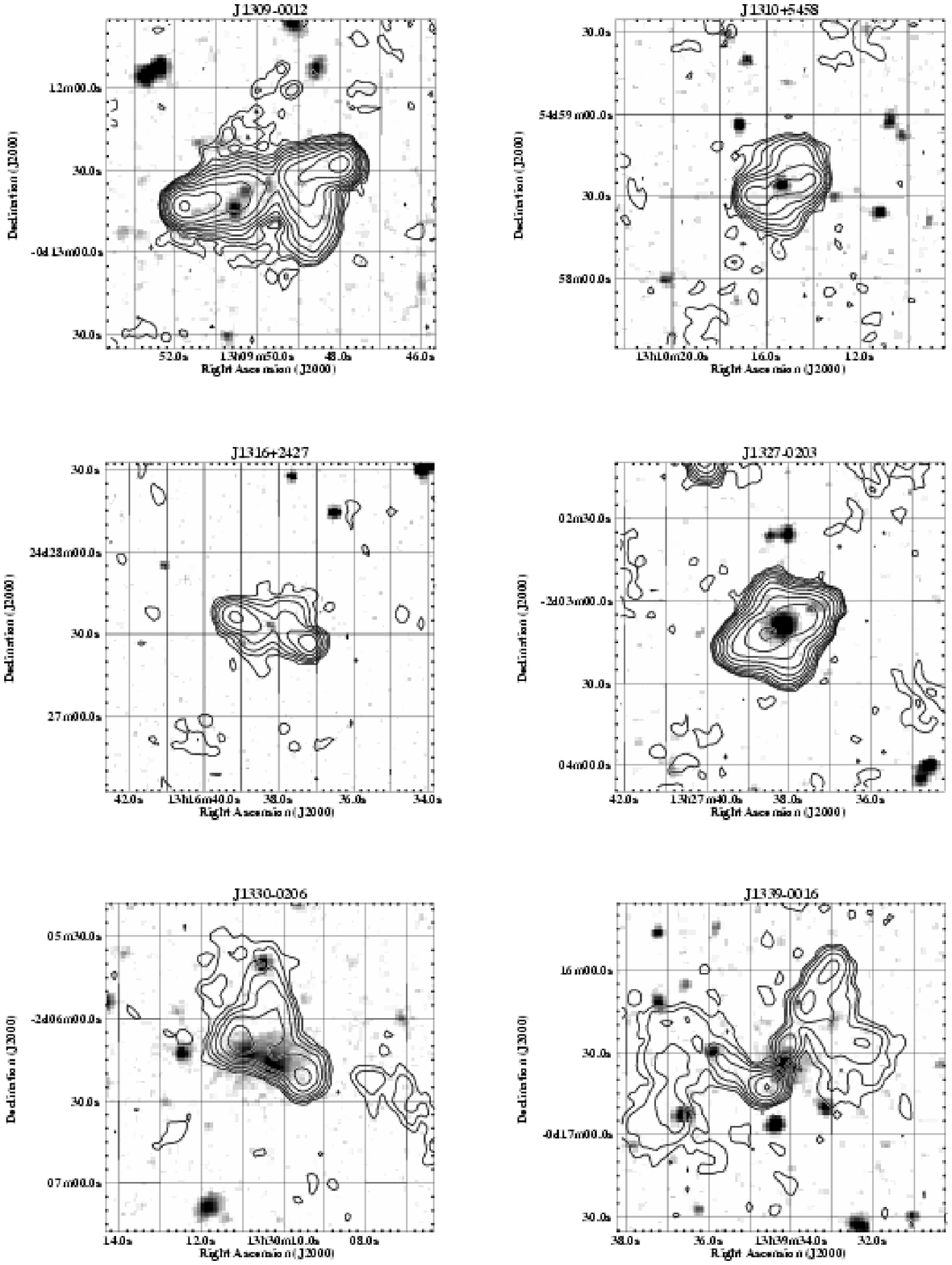} \end{figure}
\addtocounter{figure}{0}
\begin{figure} \includegraphics[width=7.0in]{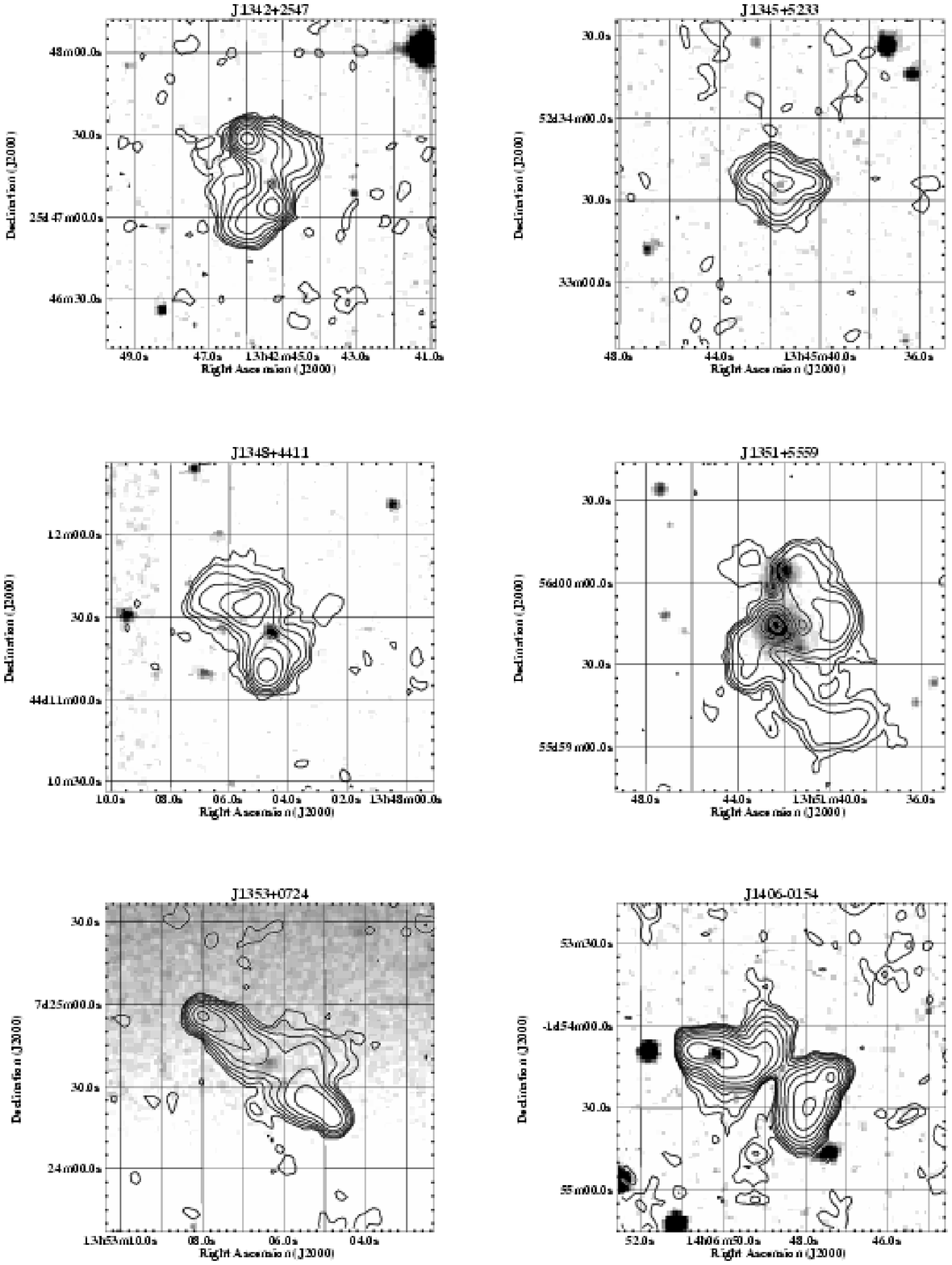} \end{figure}
\addtocounter{figure}{0}
\begin{figure} \includegraphics[width=7.0in]{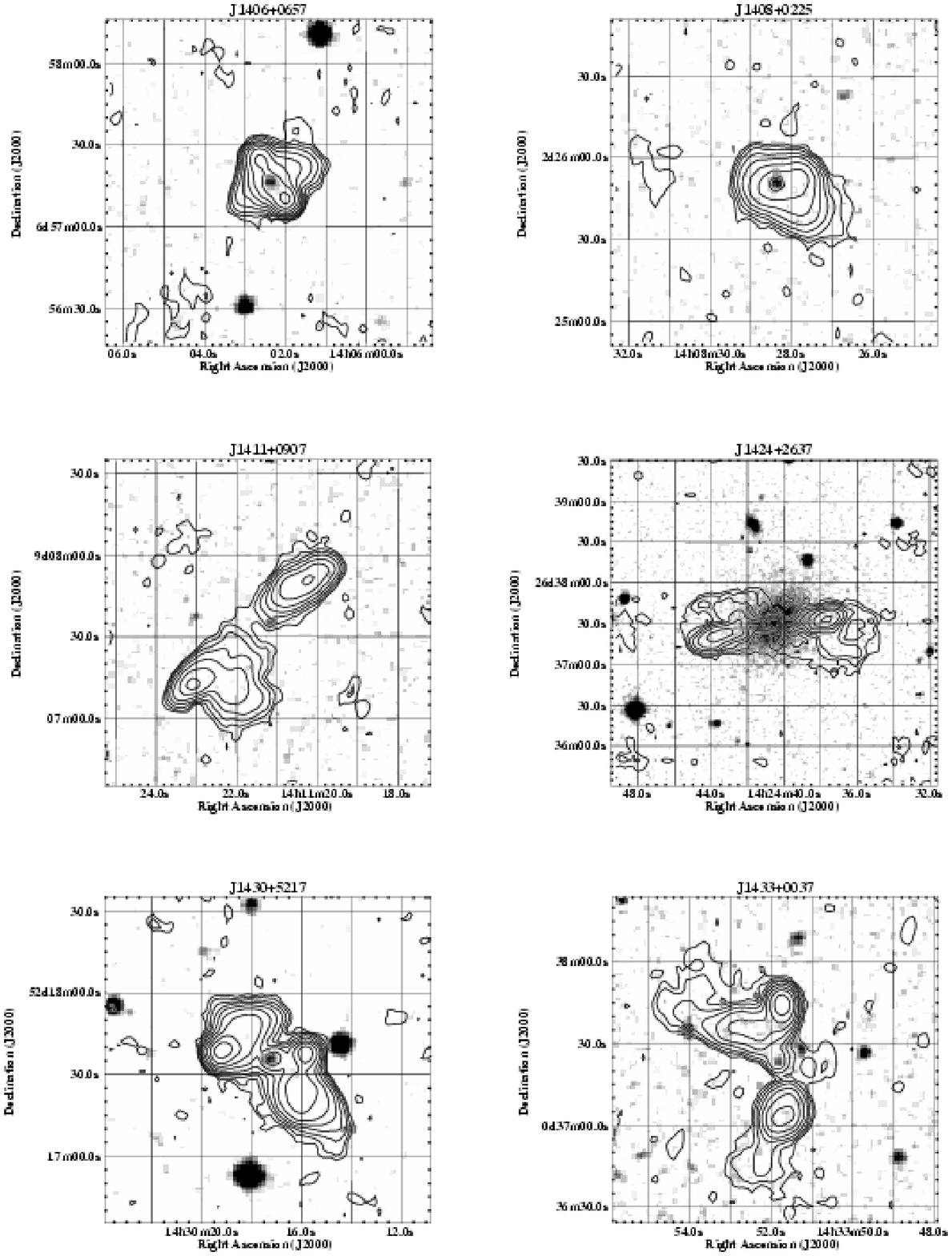} \end{figure}
\addtocounter{figure}{0}
\begin{figure} \includegraphics[width=7.0in]{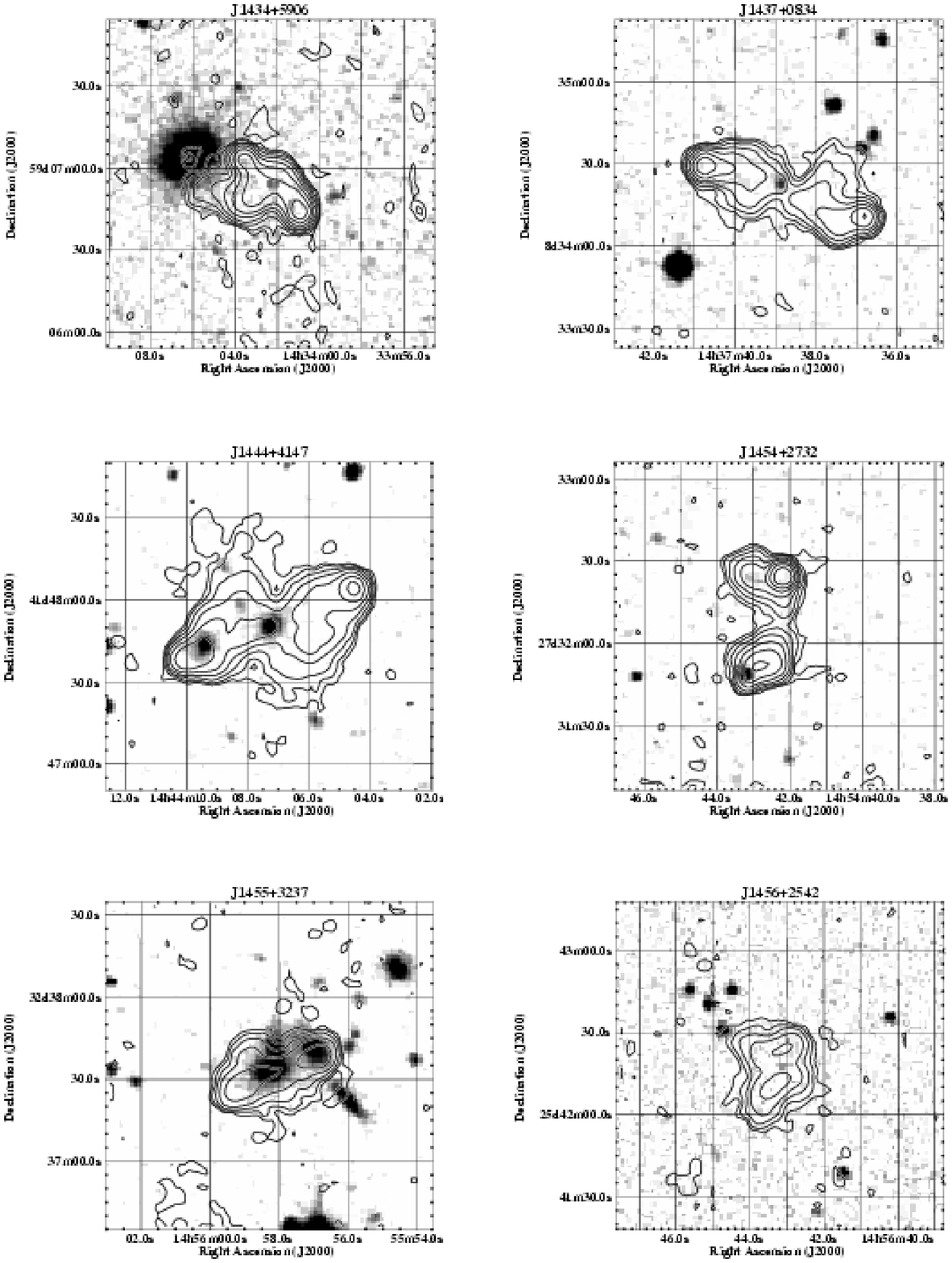} \end{figure}
\addtocounter{figure}{0}
\begin{figure} \includegraphics[width=7.0in]{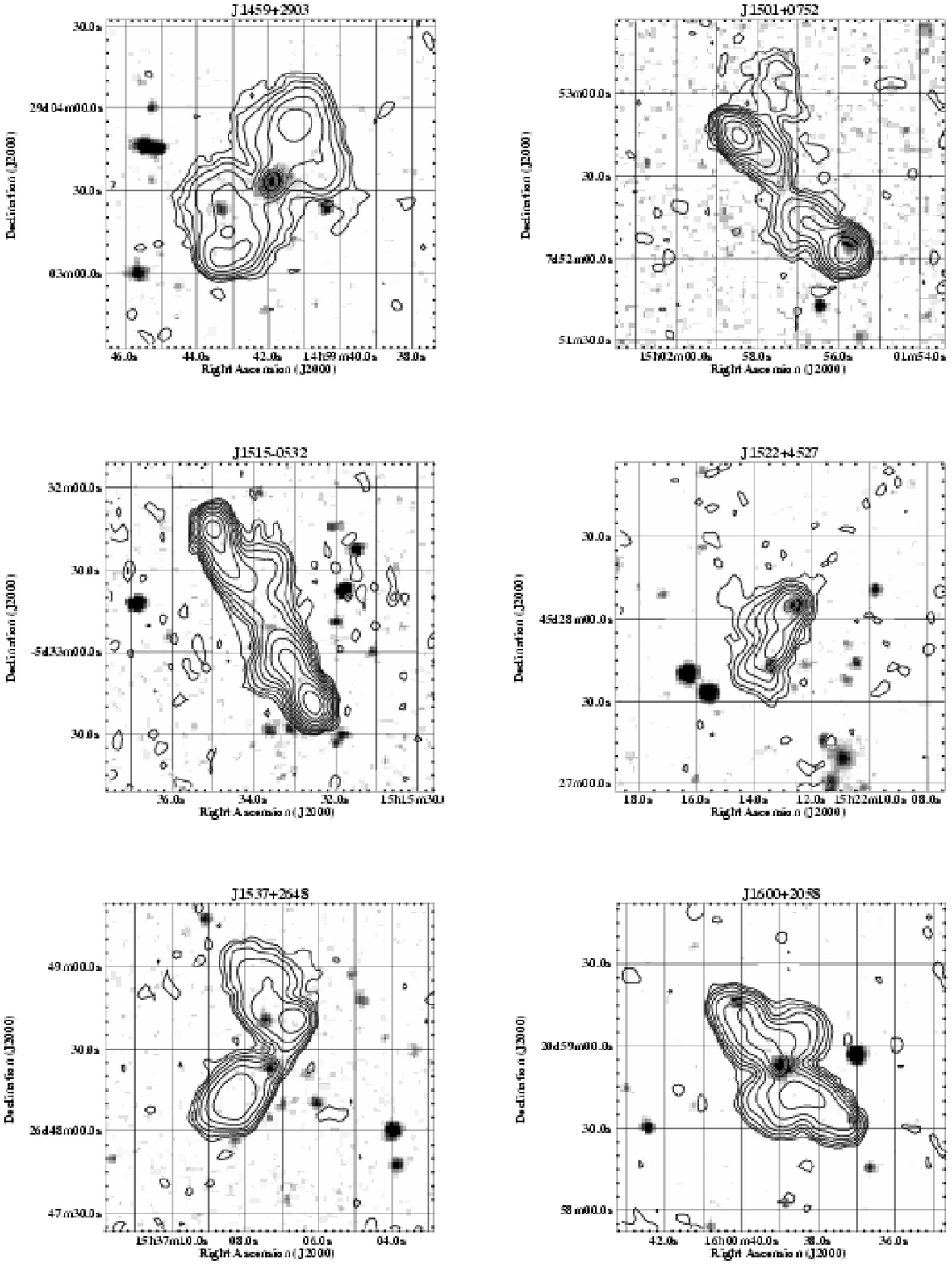} \end{figure}
\addtocounter{figure}{0}
\begin{figure} \includegraphics[width=7.0in]{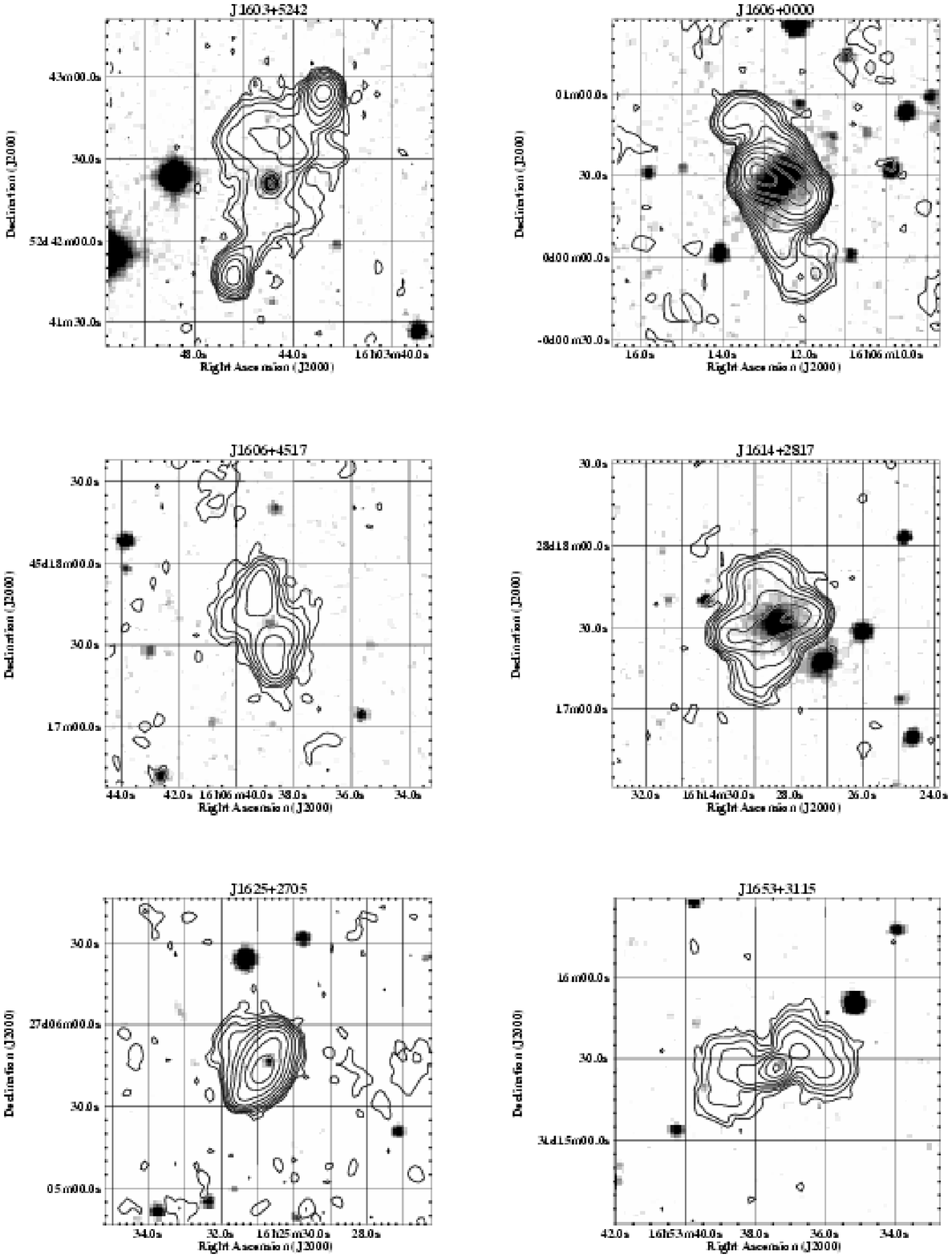} \end{figure}
\addtocounter{figure}{0}
\begin{figure} \includegraphics[width=7.0in]{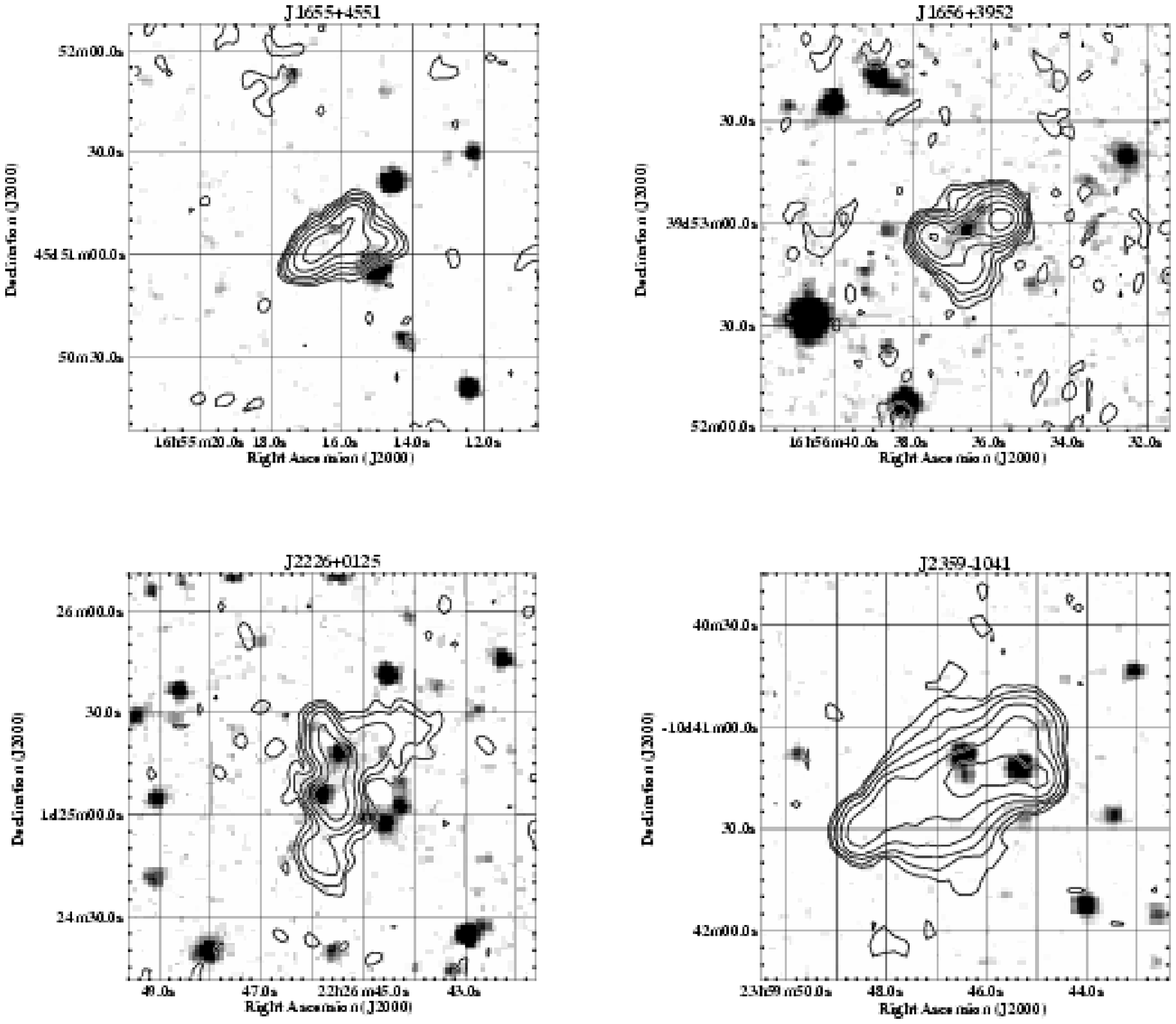} \end{figure}

\begin{figure}
\includegraphics[width=6.5in]{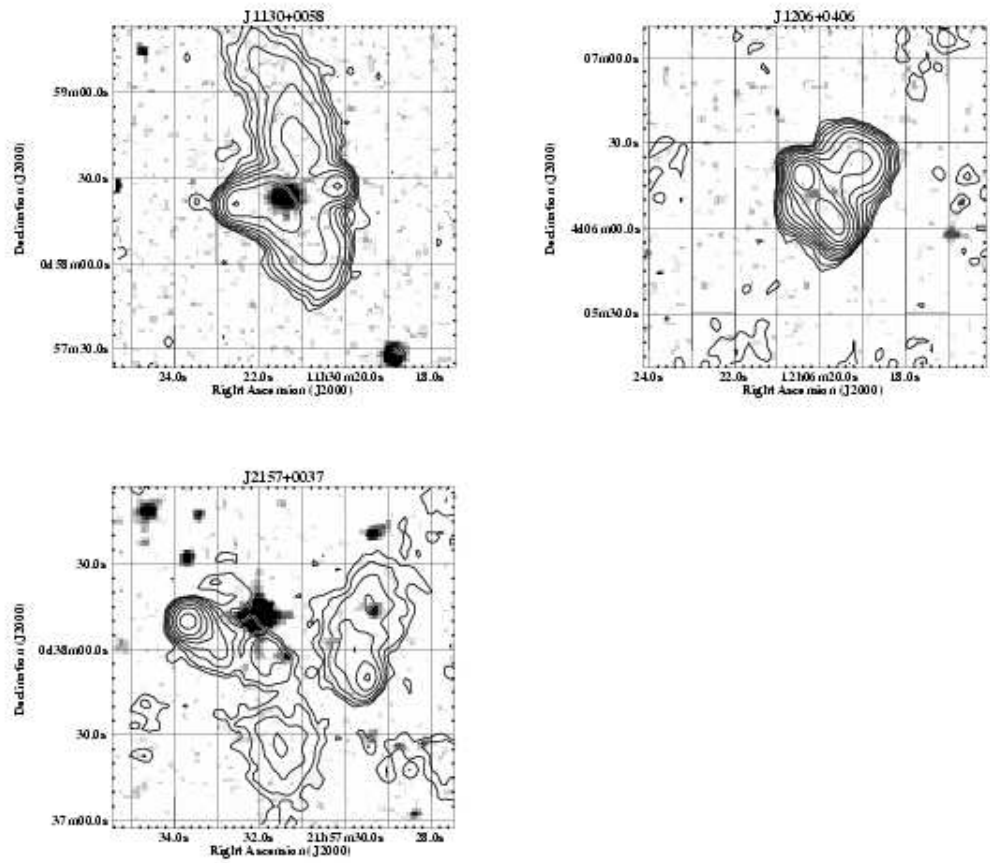}
\figcaption[f5.eps]{\label{fig-app}
As in Figure~\ref{fig-new}, but for 3 of the X-shaped sources from the 
literature described in the Appendix. } \end{figure}

\end{document}